\documentclass[12pt,onecolumn]{IEEEtran}
\usepackage[cmex10]{amsmath}
\usepackage{amsfonts}
\usepackage[pdftex]{color,graphicx}
\usepackage{subfig}
\usepackage{rotating}

\newtheorem{thm}{Theorem}

\newcommand{\defineqq}{\ensuremath{\stackrel{\text{def}}{=}}}
\newcommand{\wtilde}{\widetilde}
\newcommand{\wbar}{\overline}

\newcommand{\up}{\uparrow}
\newcommand{\down}{\downarrow}

\begin{document}

\title{Interference Channels with\\ 
Destination Cooperation}

\author{Vinod M. Prabhakaran and Pramod Viswanath\thanks{The authors are with
the Coordinated Science Laboratory, 
University of Illinois at Urbana-Champaign,
Urbana, IL 61801. Email: \{vinodmp,pramodv\}@illinois.edu}}

\maketitle

\begin{abstract}

Interference is a fundamental feature of the wireless channel. To better
understand the role of cooperation in interference management, the two-user
Gaussian interference channel where the destination nodes can cooperate by
virtue of being able to {\em both} transmit and receive is studied. The
sum-capacity of this channel is characterized up to a constant number of
bits.  The coding scheme employed builds up on the superposition scheme of
Han and Kobayashi for two-user interference channels without cooperation.
New upperbounds to the sum-capacity are also derived.

\end{abstract}

\begin{keywords}
Cooperation, interference channel, relay channel, sum-capacity.
\end{keywords}

\section{Introduction}

Orthogonalization and treating interference as noise are the two most
common ways of handling interference in practical wireless communication
systems. However, it is well-known that better rates of operation may be
achieved when the interfering systems are designed jointly as modelled in
interference channels~\cite{Carleial78,EtkinTseWang08}. Superposition
coding and interference alignment have been shown to perform well for
interference channels (where the sources only transmit and destinations
only receive)~\cite{EtkinTseWang08,CadambeJafar08}.

A further degree of cooperation is possible when the radios can both
receive and transmit. Understanding the gains from this form of cooperation
is the goal of this paper. We study a two-user Gaussian interference
channel where the destinations are also equipped with transmit
capabilities. This sets up the possibility of cooperation among the
destination nodes. The cooperative links are over the same frequency band
as the rest of the links. In this paper, we study the sum-rate when the
destination nodes operate in full-duplex. The main result is a
characterization of the sum-capacity within a constant number of bits. The
constant we obtain is 43 bits, but we discuss how this gap could be
improved. The two-user interference channel where the source radios have
receive capabilities is explored in a companion paper~\cite{SourceCoop}. A
reversibility property exists between the two
results~\cite[Section~7.1]{SourceCoop}. As we discuss there, one setting
can be viewed as being obtained from the other by (a) reversing the roles
of sources and destinations and (b) changing the directions of the links
while preserving the channel coefficients. The sum-capacities of the two
settings connected by this transformation are within a constant gap.

A scheme based on superposition coding of
Cover~\cite{CoverBroadcast} was proposed by Han and
Kobayashi~\cite{HanKobayashi} for the two-user interference channel.
Recently, Etkin, Tse, and Wang~\cite{EtkinTseWang08} showed that the scheme
of Han and Kobayashi achieves the capacity of the
two-user Gaussian interference channels to within one bit.
The scheme of Han and Kobayashi involves the two destinations partially
decoding the interference they receive. In order to facilitate this, the
sources encode their messages as a superposition of two partial messages.
One of these partial messages, termed the {\sf public} message, is decoded
by the destination where it appears as interference along with the two
partial messages which are meant for this destination. The other partial
message, called the {\sf private} message, from the interfering source is
treated as noise. Our achievable scheme employs two additional types of
messages which take advantage of cooperation:
\begin{itemize}
\item {\sf Cooperative private} messages are decoded by the destination to
which is intended, but unlike {\sf private} messages, they benefit from
cooperation. The effect of cooperation is to ensure that these messages do
not appear as interference at the destination to which they are not
intended. This is achieved using a nulling scheme.
\item {\sf Cooperative public} messages are decoded by both destinations,
and unlike {\sf public} messages, they benefit from cooperation. The form
of cooperation involves destinations exchanging messages with each other
which carry information on their past observations. This has similarities
to {\em compress-and-forward} schemes used in relay
channels~\cite{ElGamalCoverRelay}.
\end{itemize}
We also derive upperbounds on the sum-rate to show that these modes
of cooperation are optimal up to a constant gap.

Related works include~\cite{HostMadsen06,Mohajer08,Zhou10,Wang09}. The same
model was studied in~\cite{HostMadsen06}, but a constant-gap result was not
obtained there. A two-stage, two-source interference network is studied
in~\cite{Mohajer08}. Two-user Gaussian interference channels with
conferencing decoders (where the decoders communicate over an orthogonal
conferencing channel) are studied in~\cite{Zhou10,Wang09}. One-sided
interference channel with unidirectional conferencing between decoders is
considered in~\cite{Zhou10}, while~\cite{Wang09} derives the capacity
region of the two-user Gaussian interference channel with conferencing
decoders within a gap of two bits.

\section{Problem Statement}
\begin{figure*}[!t]
\centerline{
\scalebox{0.5}{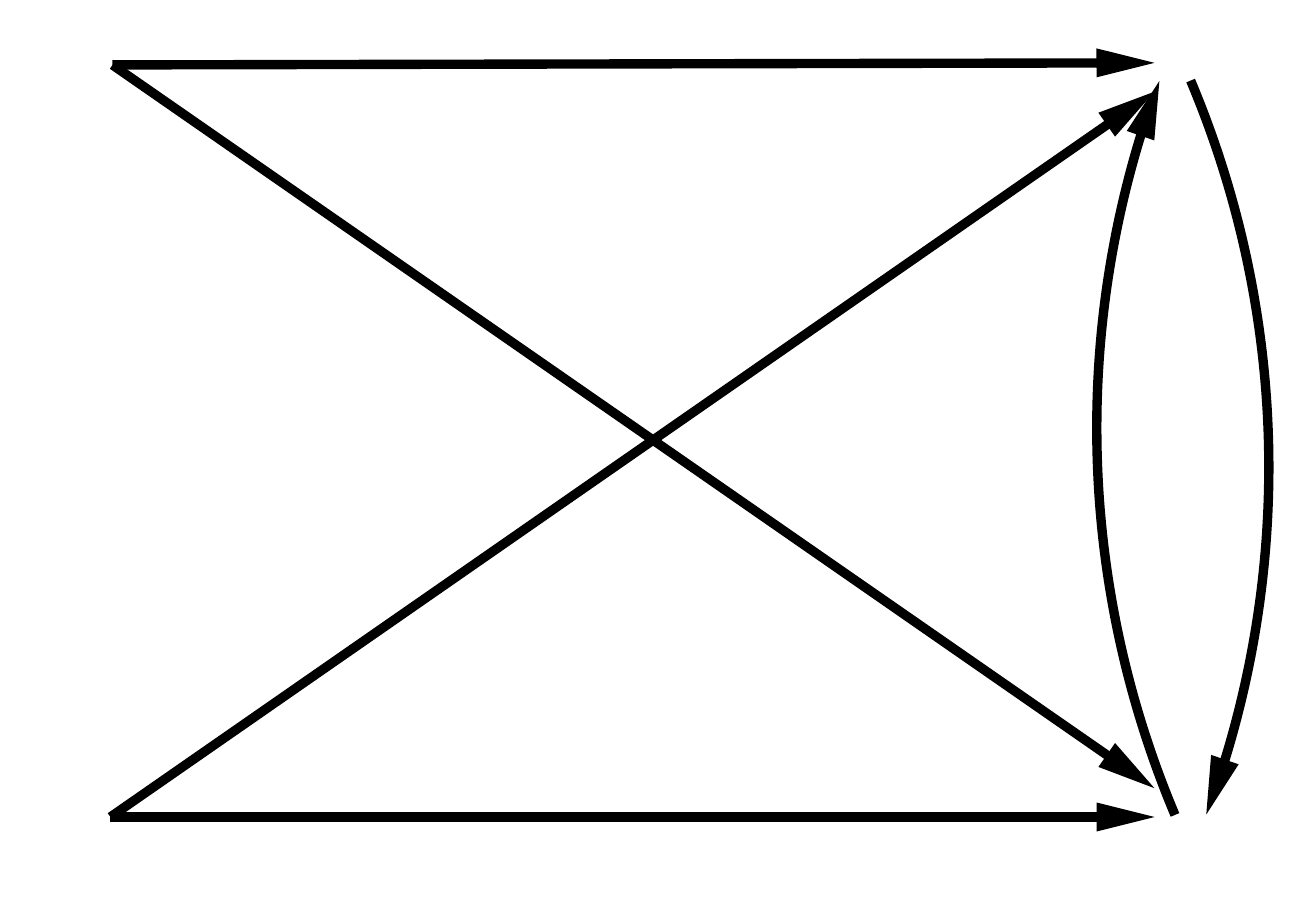}
}
\caption{Problem Setup}
\label{fig:setup}
\end{figure*}

We consider the following model for destination cooperation (see
Figure~\ref{fig:setup}). At each discrete-time instant -- indexed by
$t=1,2,\ldots$ -- the source nodes 1 and 2 send out, respectively, $X_1[t]$
and $X_2[t]$ which belong to the set ${\mathbb C}$ of complex numbers. The
destination nodes 3 and 4 can not only receive, but also transmit over the
same channel. Let $X_3[t]$ and $X_4[t]\in{\mathbb C}$, respectively, denote
what nodes 3 and 4 transmit at time $t$. Then the destination nodes receive
\begin{align*}
Y_3[t]&=g_{1,3}X_1[t]+g_{2,3}X_2[t]+g_{4,3}X_4[t]+N_3[t],\\
Y_4[t]&=g_{2,4}X_2[t]+g_{1,4}X_1[t]+g_{3,4}X_3[t]+N_4[t],
\end{align*}
where $N_3[t]$ and $N_4[t]$ are i.i.d. (over $t$), circularly symmetric,
zero-mean, unit variance, complex Gaussian random variables which are
independent of each other. The $g$'s are constant, complex channel
coefficients which are assumed to be known to all the nodes. We impose a
natural causality constraint on the transmissions from the destination nodes
-- the transmissions from each destination is a deterministic function of
everything it has received up to the previous time instant. {\em i.e.},
\begin{align*}
X_k[t]&=f_{k,t}(Y_k^{t-1}),\;k=3,4.
\end{align*}
The source nodes~1 and 2 map their messages (which are assumed to be
uniformly distributed over their alphabets and denoted by $M_1$ and $M_2$,
respectively) to their channel inputs using deterministic encoding
functions.
\begin{align*}
X_k[t]&=f_{k,t}(M_k),\;k=1,2,
\end{align*}
It is easy to see that, without loss of generality, we may consider
a channel where the channel coefficients $g_{1,3},g_{2,4},g_{3,4},g_{4,3}$ are
replaced by their magnitudes $|g_{1,3}|, |g_{2,4}|, |g_{3,4}|, |g_{4,3}|$,
and the channel coefficient $g_{1,4}$ is replaced by
$|g_{1,4}|e^{j\theta/2}$ and $g_{2,3}$ is replaced by
$|g_{2,3}|e^{j\theta/2}$, where $\theta \defineqq \arg(g_{1,4}) +
\arg(g_{2,3}) - \arg(g_{1,3}) - \arg(g_{2,4})$. We will consider this
channel. We will also assume that $|g_{3,4}|=|g_{4,3}|=g_C$, say, which
models the reciprocity of the link between nodes 3 and 4. Further, we 
will consider unit power constraints which is without loss of generality
when both destinations have the same power constraint. Thus, a
blocklength-$T$ codebook of rate $(R_1,R_2)$ is a
sequence of encoding functions, $f_{k,t},\;t=1,2,\ldots,T$ as described
above such that
\[{\mathbb E}\left[\frac{1}{T}\sum_{t=1}^T|X_k[t]|^2\right]\leq
1,\;k=1,2,3,4,\] 
with message alphabets ${\mathcal M}_k=\{1,2,\ldots,2^{TR_k}\},\;k=1,2$
over which the messages $M_k$ are uniformly distributed, and decoding
functions $\tilde{f}_{k+2}:{\mathcal C}^{T}\rightarrow{\mathcal M}_k$. We
say that a rate $(R_1,R_2)$ is achievable if there is sequence of rate
$(R_1,R_2)$ codebooks such that as $T\rightarrow \infty$,
 \[{\mathbb P}\left(\tilde{f}_{k+2}(Y_{k+2}^T)\neq M_k\right)
           \rightarrow 0,\;k=1,2.\]
In this paper, we are interested in the largest $R_1+R_2$ such that
$(R_1,R_2)$ is achievable.

We would also like to consider a linear deterministic
model~\cite{Avestimehr07} for the above channel. In order to treat both
models together, we will adopt the following notation:
The destination nodes receive
\begin{align*}
Y_3[t]&=g_{1,3}(X_1[t])+g_{2,3}^\ast(X_2[t])+g_{4,3}(X_4[t]),\\
Y_4[t]&=g_{2,4}(X_2[t])+g_{1,4}^\ast(X_1[t])+g_{3,4}(X_3[t]),
\end{align*}
where the (deterministic) encoding functions at the sources are of the form
\begin{align*}
X_k[t]&=f_{k,t}(M_k),\;k=1,2,
\end{align*}
and the (deterministic) relaying functions at the destinations are of the form
\begin{align*}
X_k[t]&=f_{k,t}(Y_k^{t-1}),\;k=3,4.
\end{align*}
\noindent{\em Gaussian case:}
\begin{align*}
g_{1,3}(X_1)&=g_{1,3}X_1,\\
g_{2,4}(X_2)&=g_{2,4}X_2,\\
g_{2,3}^\ast(X_1)&=g_{2,3}X_4+N_3,\\
g_{1,4}^\ast(X_1)&=g_{1,4}X_1+N_4,\\
g_{3,4}(X_3)&=g_{3,4}X_3,\\
g_{4,3}(X_4)&=g_{4,3}X_4,
\end{align*}
Note that $g^\ast$'s denote randomized maps while $g$'s are deterministic.

\noindent{\em Linear deterministic case:}
Let $n_{1,3},n_{2,3}$, $n_{1,4},n_{2,4}$, $n_{3,4},n_{4,3}$ be non-negative
integers and $n\defineqq\max(n_{1,3},n_{2,3}, n_{1,4},n_{2,4}
,n_{3,4},n_{4,3})$. The inputs to the channel $X_1$ and $X_2$ are
$n$-length vectors over a finite field ${\mathbb F}$. 
Let ${\bf S}$ the $n\times n$ shift matrix\footnote{In the sequel, we will
also (ab)use notation to denote ${\bf S}^0 \defineqq I$, the $n\times n$
identity matrix, and ${\bf S}^{-k} \defineqq \text{transpose}\left({\bf
S}^k\right), k\geq 0$.}.
\[{\bf S}=\left(\begin{array}{ccccc}
 0&0&0&\ldots&0\\
 1&0&0&\ldots&0\\
 0&1&0&\ldots&0\\
 \vdots&\ddots&\ddots&\ddots&\\
 0&\ldots&0&1&0
 \end{array}\right)_{n\times n}.\]
We define
\begin{align*}
g_{1,3}(X_1)&={\bf S}^{n-n_{1,3}}X_1,\\
g_{2,4}(X_2)&={\bf S}^{n-n_{2,4}}X_2,\\
g_{1,4}^\ast(X_1)&={\bf S}^{n-n_{1,4}}X_1,\\
g_{2,3}^\ast(X_2)&={\bf S}^{n-n_{2,3}}X_2,\\
g_{3,4}(X_3)&={\bf S}^{n-n_{3,4}}X_3,\\
g_{4,3}(X_4)&={\bf S}^{n-n_{4,3}}X_4.
\end{align*}
Further, to model the reciprocity of the links between the two receivers,
we set $n_{3,4}=n_{4,3}=n_C$. The set of achievable $(R_1,R_2)$ are defined
as in the Gaussian case.
\section{Main Results}
We will first state our main result on the sum-rates of the channels
presented in the previous section. Then we illustrate the gains resulting
from cooperation using an example.
\subsection{Sum-rate Characterization}

\begin{thm}{\em Linear deterministic case.}\label{thm:destcoopLD}
The sum-capacity of the linear deterministic channel with destination
cooperation is the minimum of the following
\begin{align}
u_1&=\max(n_{1,3}-n_{1,4}+n_C,n_{2,3},n_C) +
\max(n_{2,4}-n_{2,3}+n_C,n_{1,4},n_C),\label{eq:LDu1}\\
u_2&=
\max(n_{2,4},n_{2,3}) + \left(\max(n_{1,3},n_{2,3},n_C)-n_{2,3}\right),
\label{eq:LDu2}\\
u_3&=
\max(n_{1,3},n_{1,4}) + \left(\max(n_{2,4},n_{1,4},n_C)-n_{1,4}\right),
\label{eq:LDu3}\\
u_4&=\max(n_{1,3},n_C)+\max(n_{2,4},n_C),\text{ and}\label{eq:LDu4}\\
u_5&=\left\{\begin{array}{ll}
\max(n_{1,3}+n_{2,4},n_{1,4}+n_{2,3}), 
 &\text{ if }n_{1,3}-n_{2,3}\neq n_{1,4}-n_{2,4},\\
\max(n_{1,3},n_{2,4},n_{1,4},n_{2,3}),
 &\text{ otherwise}.
\end{array}\right.\label{eq:LDu5}
\end{align}
\end{thm}
The condition $n_{1,3}-n_{2,3}=n_{1,4}-n_{2,4}$ refers to a degenerate case
where
one of the destinations is degraded with respect to the other, {\em
i.e.}, the signal that one of the destinations receives is part of what the
other receives. We prove the achievability in
appendix~\ref{app:destcoopLDachievability} and the upperbound in
appendix~\ref{app:upperbounds}.

\begin{thm}{\em Gaussian case.}\label{thm:destcoopG}
The sum-capacity of the Gaussian channel with destination cooperation is at most
the minimum of the following five quantities and a sum-rate can be achieved
within a gap of at most 43 bits of this minimum\footnote{All logarithms
in this paper are to the base 2.}.
\begin{align}
u_1&=
\left\{\begin{array}{ll}
             \log\left(1 +\left(|g_{2,3}|+|g_C|+
              \left|\frac{g_{1,3}g_C}{g_{1,4}}\right|\right)^2
              +\left|\frac{g_{1,3}}{g_{1,4}}\right|^2\right), 
                &\text{ if } |g_{1,4}|>\max(1,|g_C|)\\
             \log\left(1 +\left(|g_{2,3}|+|g_C|+
              |g_{1,3}|\right)^2\right), &\text{ otherwise}
        \end{array}\right. \notag\\  
    &\quad +   \left\{\begin{array}{ll}
              \log\left(1 +\left(|g_{1,4}|+|g_C|+
              \left|\frac{g_{2,4}g_C}{g_{2,3}}\right|\right)^2
              +\left|\frac{g_{2,4}}{g_{2,3}}\right|^2\right),
                &\text{ if } |g_{2,3}|>\max(1,|g_C|)\\
             \log\left(1 +\left(|g_{1,4}|+|g_C|+
               |g_{2,4}|\right)^2\right), &\text{ otherwise,}
        \end{array}\right.\label{eq:Gaussu1}\\
u_2&=\log(1+(|g_{1,3}|+|g_{2,3}|+|g_C|)^2) +
\log\left(1+\frac{|g_{2,4}|^2}{\max(1,|g_{2,3}|^2)}\right),\label{eq:Gaussu2}\\
u_3 &=\log(1+(|g_{2,4}|+|g_{1,4}|+|g_C|)^2) +
\log\left(1+\frac{|g_{1,3}|^2}{\max(1,|g_{1,4}|^2)}\right),\label{eq:Gaussu3}\\
u_4 &= \log\left(1+(|g_{1,3}|+|g_C|)^2\right) +
    \log\left(1+(|g_{2,4}|+|g_C|)^2\right),\label{eq:Gaussu4}\\
u_5 &= \log\bigg( 1 + 
2 \left(|g_{1,3}|^2+|g_{2,4}|^2+|g_{1,4}|^2+|g_{2,3}|^2\right)\notag\\
& \qquad\qquad+ 4 \left( |g_{1,3}g_{2,4}|^2 + |g_{1,4}g_{2,3}|^2
 - 2|g_{1,3}g_{2,4}g_{1,4}g_{2,3}|\cos\theta\right) \bigg). \label{eq:Gaussu5}
\end{align}
\end{thm}
The achievability is proved in appendix~\ref{app:destcoopGachievability} and
the upperbound in appendix~\ref{app:upperbounds}.

\subsection{Gains from cooperation}

\begin{figure}
\centerline{\scalebox{0.5}{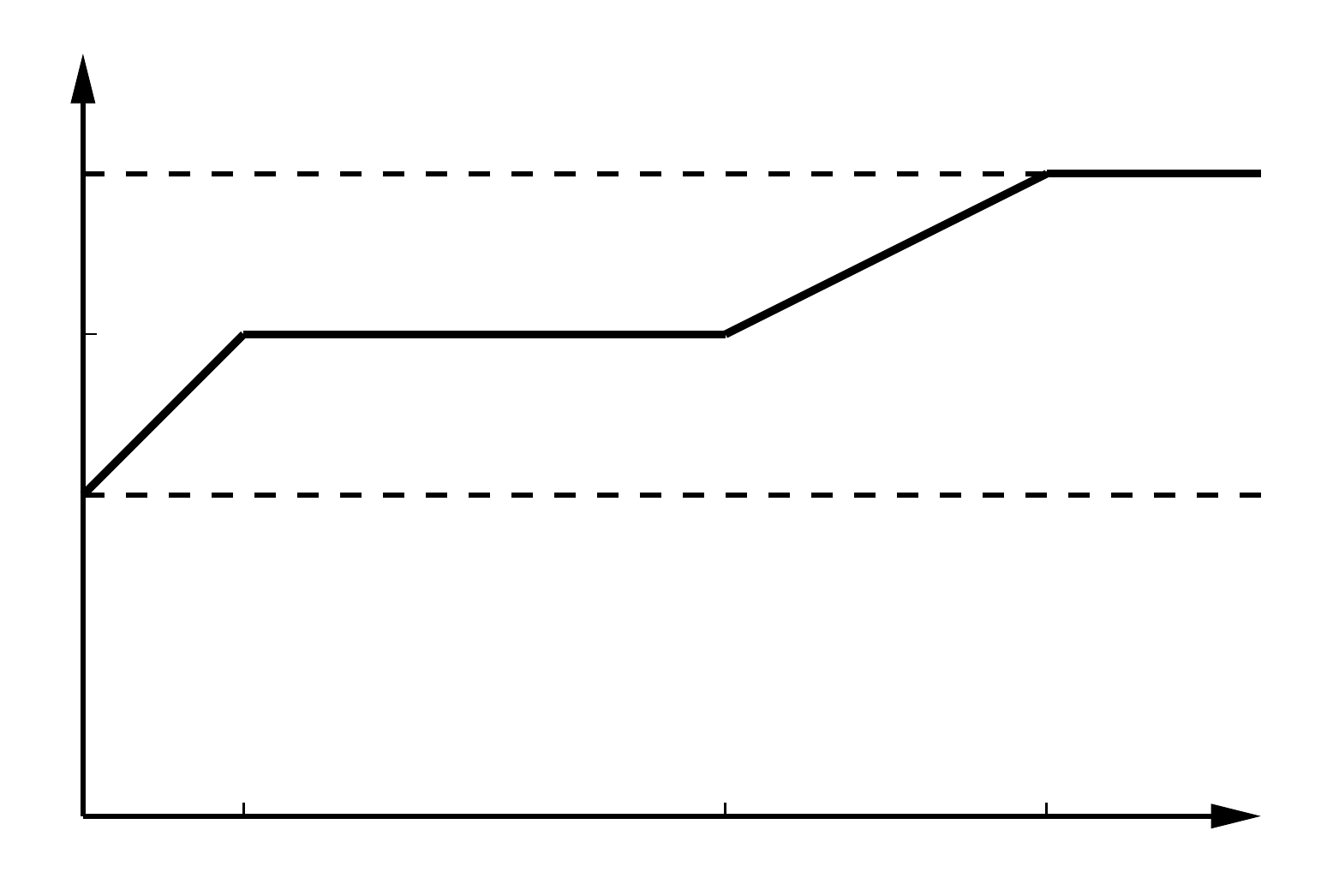}%
}
\caption{Normalized sum-capacity of the symmetric interference channel with
$g_I=\sqrt{g_D}$ under source cooperation in the limit of $g_D \rightarrow
\infty$ keeping $\log|g_I|^2/\log|g_D|^2$ and $\log|g_C|^2/\log|g_D|^2$
fixed.}
\label{fig:plot}
\end{figure}

Let us consider the following two-user Gaussian interference channel to see
the gains from cooperation: $|g_{1,3}|=|g_{2,4}|=g_D$, $|g_{1,4}| =
|g_{2,3}| = g_I = \sqrt{g_D}$, and arbitrary $\theta$. In
Fig.~\ref{fig:plot}, we plot the upperbound on the sum-rate from
Theorem~\ref{thm:destcoopG} normalized by the capacity of the direct link,
as a function of $\log |g_C|^2/\log |g_D|^2$, in the limit of
$|g_D|\rightarrow \infty$ while keeping the ratios
$\log|g_C|^2/\log|g_D|^2$ and $\log|g_I|^2/\log|g_D|^2$ constant. Since
this upperbound is achievable within a constant gap, this plot is also
that of the sum-capacity in this limit. There are three distinct regimes.
\begin{itemize}

\item $\log|g_C|^2/\log|g_D|^2 \leq 1/2.$ In this regime, the plot shows
that the capacity increases linearly with the strength of the cooperation
link (measured in the dB scale). For every 3dB increase in link strength
the sum-capacity increases by 2 bits. As will become clear in the sequel,
the benefits of cooperation in this regime come from messaging using 
{\sf cooperative private} messages.

\item $1/2 < \log|g_C|^2/\log|g_D|^2 \leq 1.$ The linear gain saturates
when the cooperation link strength is half the direct link strength. No
further gains are available until the cooperation link is as strong as the
direct link.

\item $1 < \log|g_C|^2/\log|g_D|^2 \leq 3/2.$ The capacity again increases
linearly with the cooperation link strength, but here an increase in
capacity by 2 bits requires a 6dB increase in the cooperation channel
strength. This linear increase continues until the cooperation capacity is
approached when the cooperation link is 3/2 times as strong as the direct
link, after which the capacity is flat. The cooperative gains in this
regime result from using {\sf cooperative public} messages.

\end{itemize}

\section{Coding schemes: Illustrative examples}

In this section we will present two linear deterministic channel examples
to illustrate our achievable scheme. The examples have been chosen such
that essentially signal processing schemes ({\em i.e.}, schemes which
involve no coding) can achieve the sum-capacity. Our achievable scheme for
the general problem relies on the basic intuition illustrated here.

Han and Kobayashi's achievable scheme for interference channels without
cooperation involves two kinds of signals: (a) {\sf public} signals which
are decoded by all the destinations and (b) {\sf private} signals which are
decoded only by the destinations to which they are intended and treated as
noise by all the other destinations. Our coding scheme involves two other
forms of signals. The first example introduces a third type of signal
called the {\sf cooperative private} signal. It also shows that the {\sf
private} signal itself comes in two varieties now. Example~2 introduces a
fourth type of signal called the {\sf cooperative public} signal.

\subsection{Example 1} Consider the symmetric linear deterministic
channel with direct links $n_{1,3}=n_{2,4}=n_D$, say, and interference
links $n_{1,4}=n_{2,3}=n_I$, say, such that $n_D=5$ and $n_I=2$. When
source cooperation is absent, {\em i.e.}, $n_C=0$, the sum capacity is 6.
With a cooperative link of $n_C=1$, the sum-capacity turns out to be 8. A
scheme which achieves this is as follows: The sources transmit
\[x_1[t]=\left(\begin{array}{c}
u_1[t]\\s_1[t]\\z_{\up 1}[t]\\s_1[t+1]\\z_{\down 1}[t]
\end{array}\right)\text{ and  }
x_2[t]=\left(\begin{array}{c}
u_2[t]\\s_2[t]\\z_{\up 2}[t]\\s_2[t+1]\\z_{\down 2}[t]
\end{array}\right),
\]
where $s_k(1)=s_k(T+1)=0,\;k=1,2$. The destination nodes will receive
the following signals
\begin{align*}
y_3[t]&=\left(\begin{array}{c}
u_1[t]\\s_1[t]\\z_{\up 1}[t]\\s_1[t+1]\\z_{\down 1}[t]
\end{array}\right)+
\left(\begin{array}{c}
0\\0\\0 \\u_2[t]\\s_2[t]
\end{array}\right)+
\left(\begin{array}{c}
0\\0\\0 \\0 \\-(s_2[t]+u_1[t-1])
\end{array}\right)\\
&=\left(\begin{array}{c}
u_1[t]\\s_1[t]\\z_{\up 1}[t]\\s_1[t+1]+u_2[t]\\z_{\down 1}[t]-u_1[t-1]
\end{array}\right), \text{ and similarly}\\
y_4[t]&=\left(\begin{array}{c}
u_2[t]\\s_2[t]\\z_{\up 2}[t]\\s_2[t+1]+u_1[t]\\z_{\down 2}[t]-u_2[t-1]
\end{array}\right), 
\end{align*}
if they transmit at time $t$
\[
x_3[t]=\left(\begin{array}{c}
-(s_1[t]+u_2[t-1])\\-(z_{\down 1}[t-1]-u_1[t-1])\\0\\0\\0\end{array}\right)\text{ and  }
x_4[t]=\left(\begin{array}{c}
-(s_2[t]+u_1[t-1])\\-(z_{\down 2}[t-1]-u_2[t-1])\\0\\0\\0\end{array}\right).\]
Clearly, the destinations nodes may transmit the above signals in this
example and destination~3 can recover the signals $\{(u_1[t],s_1[t],z_{\up
1}[t],z_{\down 1}[t],,u_2[t]): t=1,2,\ldots T\}$ while destination~4 can
recover the signals $\{u_2[t],s_2[t],z_{\up 2}[t], z_{\down 2}[t], u_1[t]:
t=1,2,\ldots T\}$. This implies that a rate $R_1=4$ (and similarly $R_2=4$)
is achievable.

But we now interpret the steps involved at the destinations with an
additional restriction that they can access lower levels of their
observations only if the signals contributing to the higher levels have
been recovered. As we will see in the next section, this restriction allows
us to extend this scheme to a more general scheme which also works in the
Gaussian case. The rough intuition is that the scheme will treat the
signals in the lower levels, which represent lower power levels in the
Gaussian context, as noise while processing the higher levels. And if a
higher level has not been decoded, then the lower levels are essentially
``drowned out'' by the higher power of the undecoded signals.

At the end of time $t$, destination~3 performs a preliminary decoding ({\em
phase-1 decoding}) where it recovers $u_1[t]$, $s_1[t]$, $z_{\up 1}[t]$ in
that order reading them off from the top levels of $y_3[t]$. Then, it
removes the effect of these signals from $y_3[t]$ to obtain the residual
signal
\[ \tilde{y}_3[t]= \left(\begin{array}{c}
   0\\0\\0\\s_1[t+1]+u_2[t]\\z_{\down 1}[t]-u_1[t-1]
 \end{array}\right).\]
This residual signal is multiplied by -1 and shifted upwards (equivalent to
a scaling in the Gaussian case) and transmitted at time $t+1$ as
$x_3[t+1]$.  Note that destination~3 at this point has not decoded all the
signals from $y[t]$. But phase-1 decoding of $y_3[t]$ allows it to
construct $x_3[t+1]$ at the end of time $t$. Destination~3 has to wait
till at least after time $t+1$ in order to recover $u_2[t]$ from the fourth
level of its observation at time $t$, namely $s_1[t+1]+u_2[t]$. This is
because neither $u_2[t]$ nor $s_1[t+1]$ is available separately until time
$t+1$ when the latter is recovered. The destination then removes the effect
of this level. This allows it to recover $z_{\down 1}[t]$ since $u_1[t-1]$
is already recovered. Recovering In generalizing this example, our
achievable scheme will adopt a similar approach as explained in the next
section.

\begin{sidewaysfigure}
\centering
\subfloat[]{\scalebox{0.55}{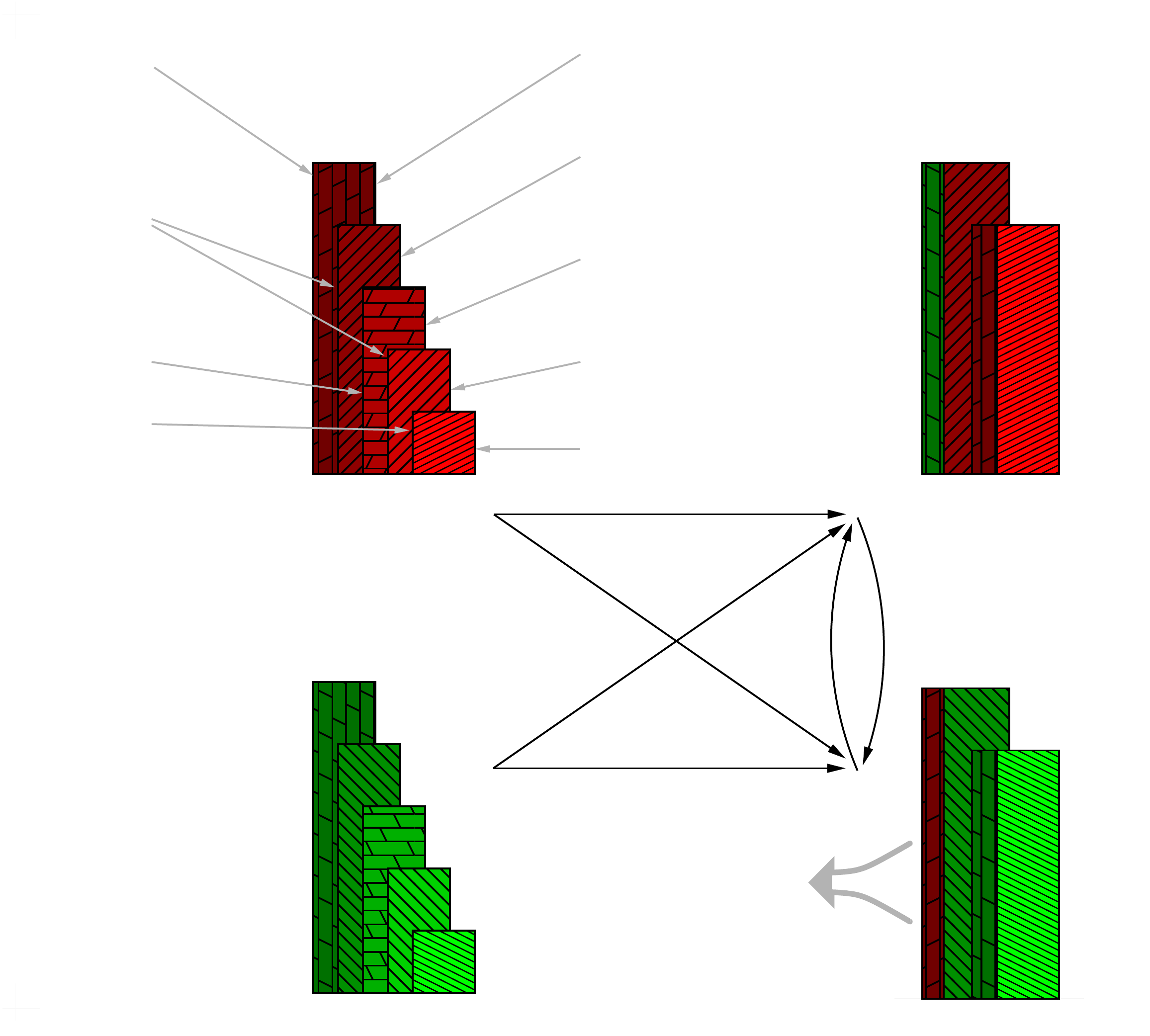}}%
\subfloat[]{\scalebox{0.55}{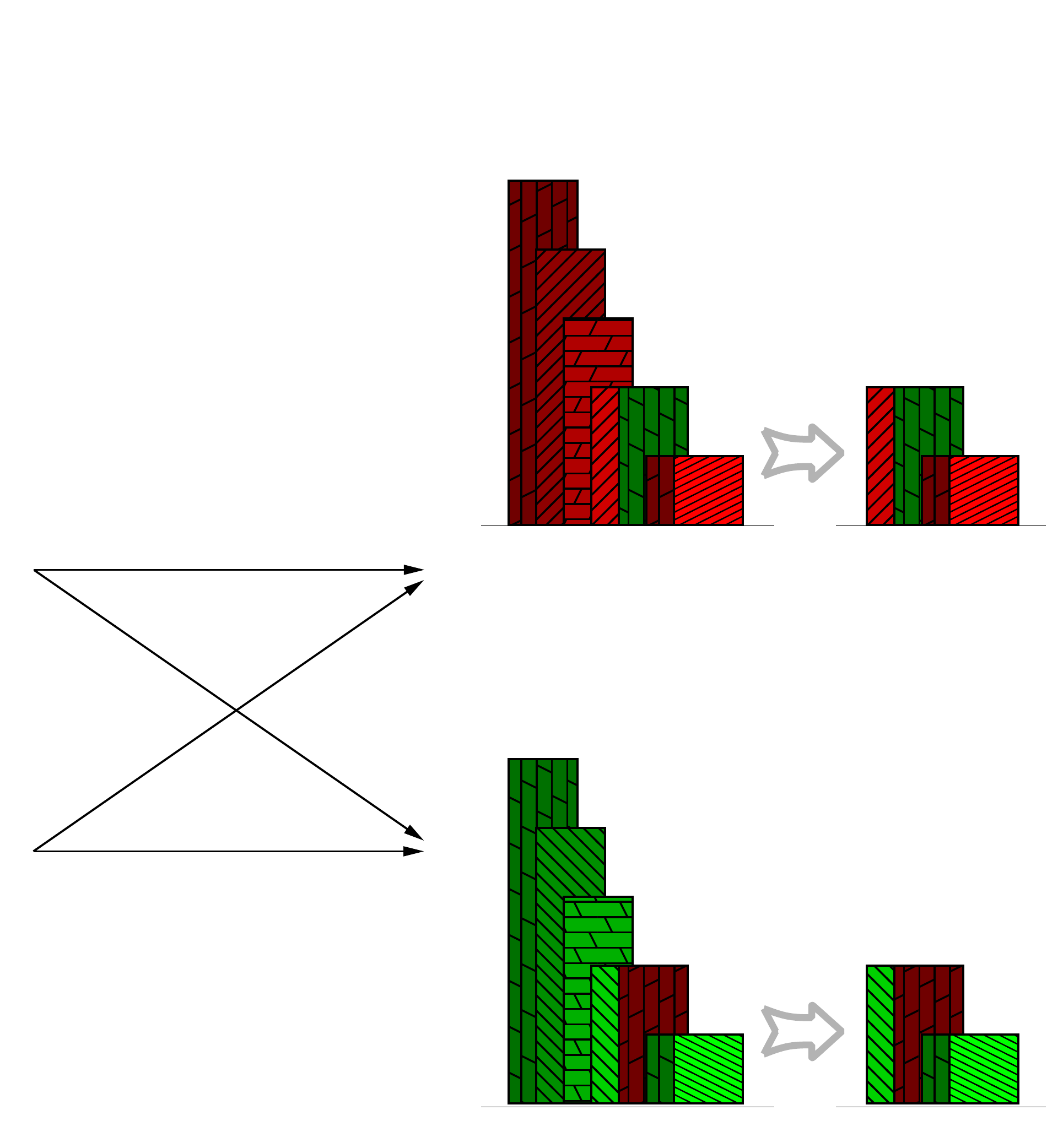}}
\caption{Example 1. (a) Transmitted signals. (b) Received signals and
residual signals after phase-1 decoding. Notice that the destinations
transmit at time $t$ a scaled up version of their residual signals from
time $t-1$. This results in a nulling-out of the interference caused by the
{\sf cooperative private} signal from the sources.}
\label{fig:example1}
\end{sidewaysfigure}

The above scheme involved three types of signals
(Figure~\ref{fig:example1}).
\begin{itemize}
\item {\sf Public} signals. $u_1$ and $u_2$ in the scheme above are
recovered by both the destinations. This signal is similar to the {\sf
public} message of Han and Kobayashi's superposition scheme for two-user
interference channels.
\item {\sf Private} signals. $z_{\up 1}$, $z_{\down 1}$, $z_{\up 2}$, and
$z_{\down 2}$ are recovered only by the destinations to which they are
intended. This signal is again similar to the {\sf private} message of Han
and Kobayashi's scheme. We want to further divide these signals into two
types in the context of our scheme:
\begin{itemize}
\item $\up$ {\sf private} signals. $z_{\up 1}[t]$ and $z_{\up 2}[t]$ are
recovered by the respective destinations on observing $y_3[t]$ and $y_4[t]$
respectively. The destinations remove the effects of these signals from the
observations in order to prepare the residual signal which is transmitted
at time $t+1$. In this sense these signals are treated differently by the
destinations compared to the signals $z_{\down 1}[t]$ and $z_{\down 2}[t]$.
\item $\down$ {\sf private} signals. $z_{\down 1}[t]$ and $z_{\down 2}[t]$,
unlike the above $\up$~{\sf private} signals, are recovered by the
respective destinations with a certain time-lag.
\end{itemize}
\item {\sf Cooperative private} signals. $s_1$ and $s_2$ are also recovered
only by the destinations to which they are intended. However, their effect
at the destination where they could act as interference is nulled out by
the actions of the other destination. In this sense, these signals benefit
cooperation.
\end{itemize}

In this example, sources made use of the relaying capabilities of
destination nodes to beamform and null-out part of the interference. This
idea has similarities to a technique independently arrived at
in~\cite{Mohajer08} which the authors call {\em interference
neutralization}.

\subsection{Example 2} Consider the symmetric linear deterministic
channel with $n_D=2, n_I=1, n_C=3$.
{\small\[
x_1[t]=\left(\begin{array}{c}
u_1[t]\\z_1[t]\\0
\end{array}\right)\text{ and }
x_2[t]=\left(\begin{array}{c}
u_2[t]\\z_2[t]\\0
\end{array}\right),\]}
where $u_k(T)=u_k(0)=0,\;k=1,2$.  The destinations transmit
{\small\[
x_3[t]=\left(\begin{array}{c}
u_1[t-1]\\0\\0
\end{array}\right)\text{ and }
x_4[t]=\left(\begin{array}{c}
u_2[t-1]\\0\\0
\end{array}\right)\]}
which is possible since the destinations receive
{\small\[
y_3[t]=\left(\begin{array}{c}
u_2[t-1]\\u_1[t]\\z_1[t]+u_2[t]
\end{array}\right)\text{ and }
y_4[t]=\left(\begin{array}{c}
u_1[t-1]\\u_2[t]\\z_2[t]+u_1[t]
\end{array}\right).\]}
Thus, destination~3 now recovers
$\{(u_1[t],z_1[t],u_2[t]):t=1,2,\ldots,T\}$ and destination~4
$\{u_2[t],z_2[t],u_1[t]:t=1,2,\ldots,T\}$ leading to rates $R_1=2,R_2=2$.
Here the destinations helped each other decode part of the interference.

\begin{sidewaysfigure}
\centering
\subfloat[]{\scalebox{0.55}{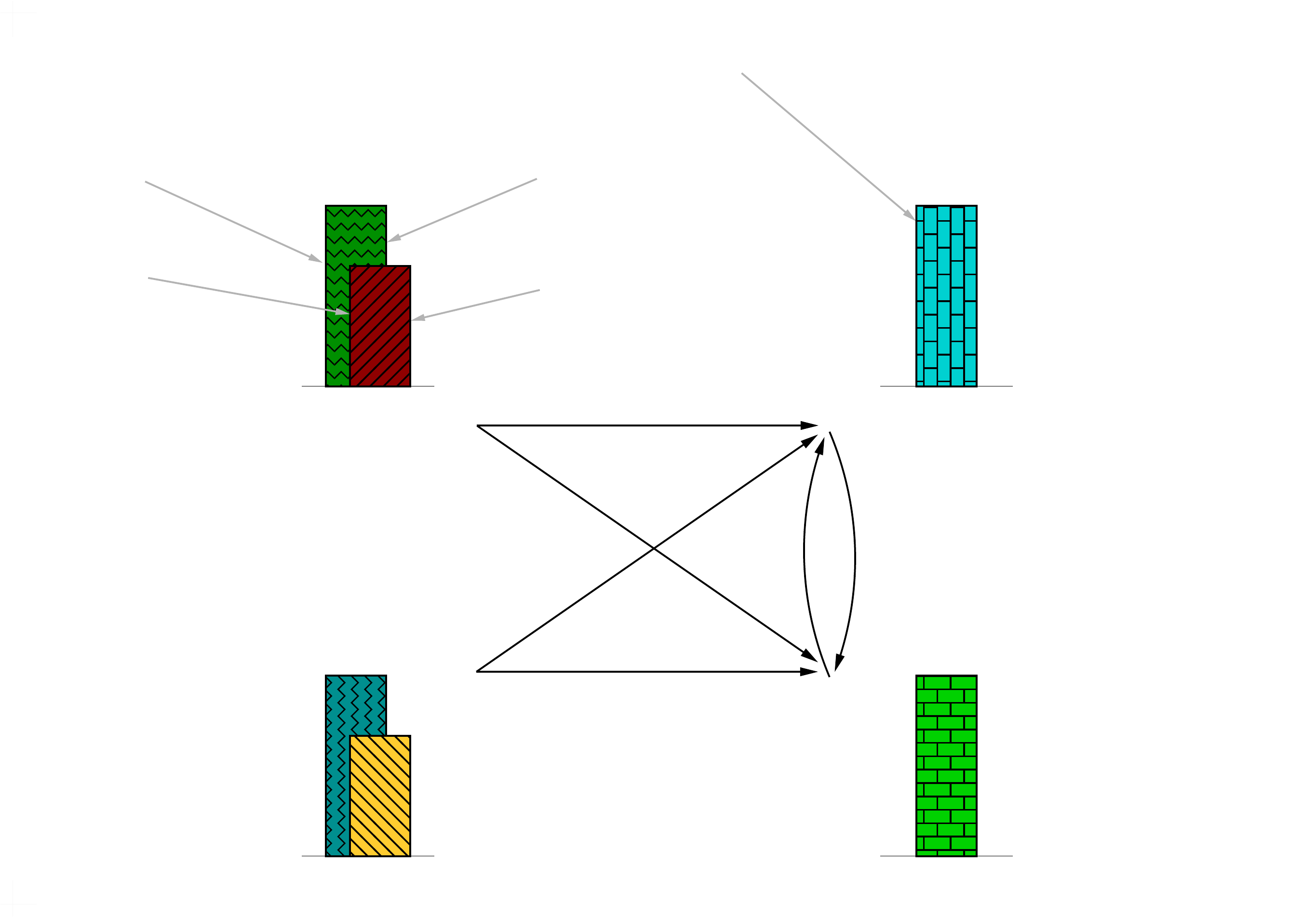}}%
\subfloat[]{\scalebox{0.55}{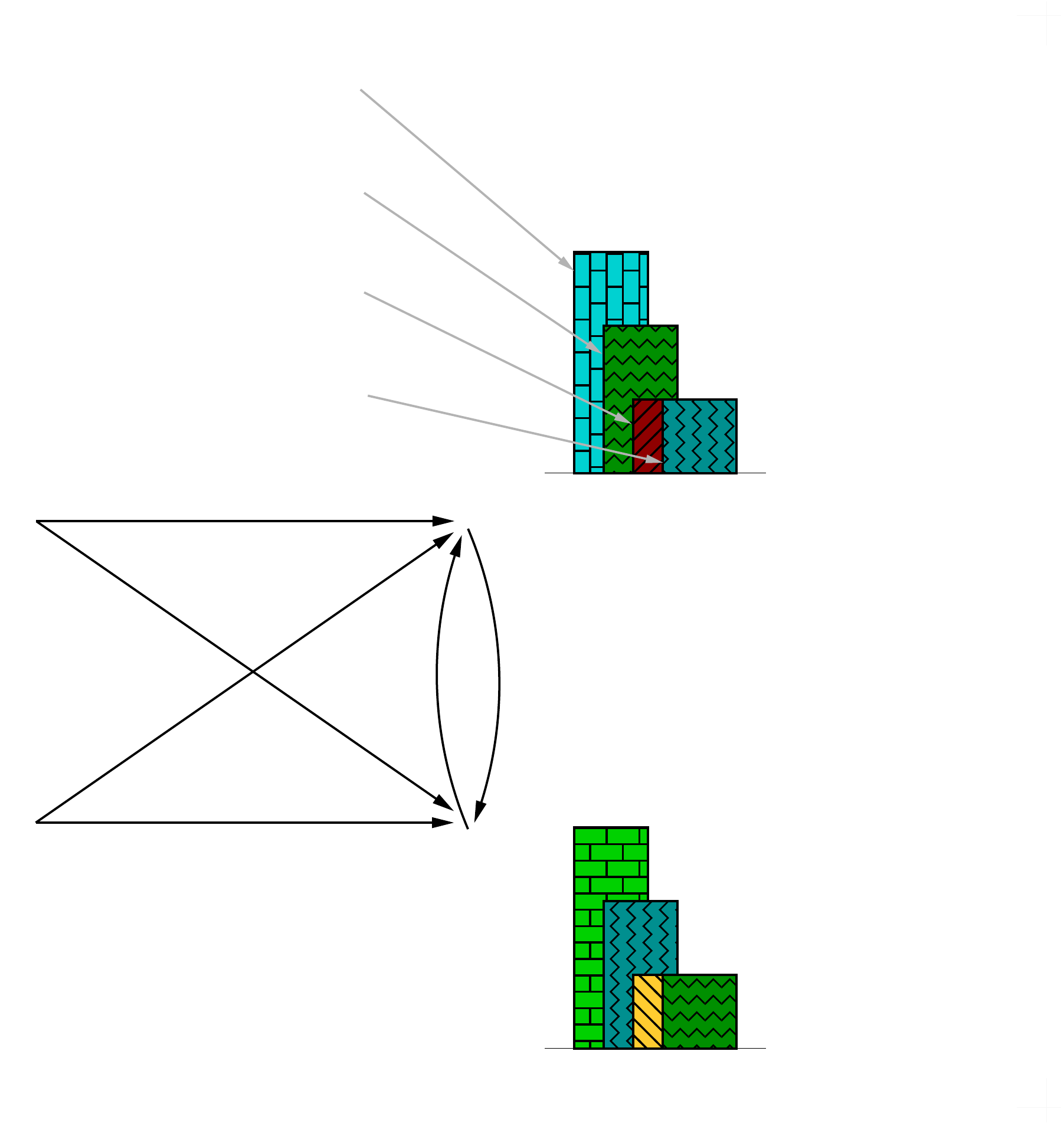}}
\caption{Example 2. (a) Transmitted signals. (b) Received signals.}
\label{fig:example2}
\end{sidewaysfigure}

We have two types of signals in this example (Figure~\ref{fig:example2}).
Signals $z_1$ and $z_2$ are {\sf private} signals in that they are decoded
only by the destination to which it is intended and they did not benefit
from cooperation. Signals $u_1$ and $u_2$ constitute a new kind of signal.
\begin{itemize}
\item {\sf Cooperative public} signals. These signals are decoded by both
destinations. Their transmission benefited from cooperation. In this
example, destination~3 aided destination~4 in recovering the signal $u_1$.
It should be noted that, if destination~3 at time $t$, instead of sending
$u_1[t-1]$, sent an appropriate linear projection of its received vector
truncated at the top two levels (so as not include any of the {\sf private}
signal in this linear projection), the scheme would continue to work. In
generalizing this example, our achievable scheme will adopt a similar
approach.
\end{itemize}

\section{Coding schemes}

We use two coding schemes to prove the achievability of our main results.
These schemes cater to different regimes of the cooperation link.  When the
cooperation link is weaker than the other links, we use a scheme which
extends the intuition from Example~1 of the previous section. The basic
intuition is that sources may make use of the relaying capabilities of
destination nodes to beamform and null-out part of the interference. When
the cooperation link is stronger than the direct links, we use a coding
scheme similar to Example~2.  This is a form of compress-and-forward scheme
where the destinations quantize their observations and convey these to each
other over the cooperation link. We roughly sketch the schemes in this
section and leave the formal proofs to the appendices. In the appendix
(\ref{app:destcoopLDachievability} and \ref{app:destcoopGachievability}),
we also show that achievability in these two regimes imply achievability
for the entire range of cooperative link strengths.

\subsection{Cooperation link weaker than other links} 

This is a block-Markov coding scheme which generalizes Example~1. The exact
details are provided in Appendices~\ref{app:destcoopLDachievability}
and~\ref{app:destcoopGachievability}, we only provide a rough sketch here
in the context of the Gaussian channel.  For each block-$j$, the different
types of messages, namely, {\sf public}, $\up$~{\sf private}, $\down$~{\sf
private}, and {\sf cooperative-private} messages are coded using
independent codebooks. For source-$k$, where $k=1,2$, let us denote these
codewords, respectively, by $c_{u_k}^{(j)}$, $c_{\up z_k}^{(j)}$, $c_{\down
z_k}^{(j)}$, and $c_{s_k}^{(j)}$, respectively. Source-$k$ transmits a
superposition of the $c_{u_k}^{(j)}$, $c_{\up z_k}^{(j)}$, $c_{\down
z_k}^{(j)}$ codewords and a signal $f_{s_k}^{(j)}$ (defined below) which
depends on the current and future $c_{u_k}$ codewords, {\em i.e.}, on
$\left\{c_{u_k}^{(i)}:\;i=j,j+1,\ldots,J\right\},$ where $J$ is the total
number of blocks.
\begin{align*}
f_{s_k}^{(j)}\defineqq
c_{s_k}^{(j)} + A_k c_{s_k}^{(j+1)} + A_k^2 c_{s_k}^{(j+2)} +
  A_k^{J-j} c_{s_k}^{(J)},
\end{align*}
where $A_k$ will be defined later. Note that $f_{s_k}$ can be thought
of as the effect of passing the signal $c_{s_k}$ through an anti-causal
filter (which acts on blocks rather than individual samples) with transfer
function
\[ \frac{1}{1-A_kz^{-1}}.\]
Destinations decode in two phases. At the end of the $j$-th block,
destination~3 decodes using successive cancellation decoding, the codewords
$c_{u_1}^{(j)}$, $c_{s_1}^{(j)}$, and $c_{\up z_1}^{(j)}$ in that order
while treating all the other undecoded codewords and interference as noise.
This constitutes the first phase of decoding. The residual signal from
which the contribution of the decoded signals is removed is scaled by a
factor $A_3$ and transmitted as the $X_3$ signal in block-$j+1$.
Destination~4 also performs its first phase of decoding in a similar manner
and obtains its transmit signal $X_4$ for block-$j+1$. Assuming that all
the decoding in previous blocks were successful, the signal at
destination~3 in block-$j$ is a linear combination of signals involving
$\{c_{u_1}^{(i)}:\; i=j-1,j\}$, $\{c_{u_2}^{(j)}\}$,
$\{c_{s_1}^{(i)}:\;i=j-1,j,\ldots,J\}$,
$\{c_{s_2}^{(i)}:\;i=j,j+1,\ldots,J\}$, $\{c_{z_{\up
1}}^{(i)}:\;i=j-1,j\}$, $\{c_{z_{\down 1}}^{(i)}:\;i=1,2,\ldots,j\}$,
$\{c_{z_{\up 2}}^{(j)}\}$, and $\{c_{z_{\down
2}}^{(i)}:\;i=1,2,\ldots,j\}$. Of these, let us consider the
contribution of $c_{s_2}$. We first define the notation $\{ f \}_{j-1}$ to 
denote $f$ from which all terms which depend on
$\{c_{s_2}^{(i)}:\;i=1,2,\ldots,j-1\}$ have been removed. For instance, for
$i\geq 0$,
\begin{align*}
 \left\{ f_{s_2}^{(j-i)} \right\}_{j-1} &= 
 \left\{ c_{s_2}^{(j-i)} + A_2 c_{s_2}^{(j-i+1)} + A_2^2
c_{s_2}^{(j-i+2)} + \ldots + A_2^{J-j+i}c_{s_2}^{(J)} \right\}_{j-1}\\
 &= A_2^ic_{s_2}^{(j)} + A_2^{i+1}c_{s_2}^{(j+1)} + \ldots +
A_2^{J-j+i}c_{s_2}^{(J)}\\&= A_2^i f_{s_2}^{(j)}.
\end{align*}
Then, the contribution of $c_{s_2}$ in the signal received at destination~3
in block-$j$ is\footnote{Note that there are only a finite number of terms
since the block indices start from $j=1$.}
\begin{align*}
g_{2,3}& f_{s_2}^{(j)} + g_C A_4 \bigg\{g_{2,4}f_{s_2}^{(j-1)}
+ g_CA_3 \left(g_{2,3}f_{s_2}^{(j-2)} + g_C A_4
g_{2,4}f_{s_2}^{(j-3)}\right)\\ 
&\qquad\qquad\qquad\qquad\qquad\qquad\quad
+ g_CA_3g_CA_4g_CA_3 \left(g_{2,3}f_{s_2}^{(j-4)} 
+ g_C A_4 g_{2,4} f_{s_2}^{(j-5)}\right) + \ldots  \bigg\}_{j-1}\\
=& \left(g_{2,3} f_{s_2}^{(j)} 
    + g_CA_4g_{2,4}\left\{f_{s_2}^{(j-1)}\right\}_{j-1}\right)
   +(g_cA_4g_CA_3)\left(g_{2,3}\left\{f_{s_2}^{(j-2)}\right\}_{j-1} +
   g_C A_4 g_{2,4}\left\{f_{s_2}^{(j-3)}\right\}_{j-1}\right)\\
&\qquad\qquad\qquad\qquad\qquad
   +(g_CA_4g_CA_3)^2\left(g_{2,3} \left\{f_{s_2}^{(j-4)}\right\}_{j-1} +
     g_C A_4 g_{2,4}\left\{f_{s_2}^{(j-5)}\right\}_{j-1}\right) + \ldots\\
=& \left(g_{2,3} + g_CA_4g_{2,4}A_2\right)
\left(1+  (g_cA_4g_cA_3)A_2^2 + (g_cA_4g_cA_3)^2A_2^4 + \ldots\right)
f_{s_2}^{(j)}.
\end{align*}
Hence, if we choose $A_2,A_4$ such that
\[ g_{2,3} = - g_C g_{2,4} A_4 A_2,\]
we can ensure that $c_{s_2}$ causes no interference at destination~3.
Similarly, choosing $A_1,A_3$ to satisfy
\[ g_{1,4} = - g_C g_{1,3} A_3 A_1\]
ensures that destination~4 receives no interference from $c_{s_1}$. In
appendix~\ref{app:destcoopGachievability} we show that there are choices for
$A$'s such that these conditions and the power constraints at the
transmitters can be satisfied in the regime of interest. The upshot of this
is that the {\sf cooperative private} messages do not cause interference at
the destinations. Appendix~\ref{app:destcoopLDachievability} employs a
similar scheme in the context of linear deterministic channels.
  
The second phase of decoding is performed after the entire transmission is
completed. At the end of the transmissions, the destinations have decoded
their own {\sf public}, {\sf cooperative private}, and $\up$~{\sf private}
messages for all blocks. They may cancel the effects of these messages from
their observations. From the residual signal they successively decode the
interferer's {\sf public} message and their own $\down$~{\sf private}
message in that order while treating the remaining interference as noise.
The power allocations and the rates calculation are deferred to he
appendices~\ref{app:destcoopLDachievability}
and~\ref{app:destcoopGachievability}.

\subsection{Cooperation link stronger than direct links} The following
theorem generalizes the scheme in Example~2. The coding scheme we use is a
block-Markov coding scheme which has elements of compress-and-forward
coding~\cite{ElGamalCoverRelay}, which was originally proposed for relay
channels, and superposition coding for interference
channels~\cite{HanKobayashi}. The destinations perform a form of backwards
decoding~\cite{WillemsBackwards}.

\begin{thm}\label{thm:destcoopgenericschemes}
Given joint distributions $p_{U_1,X_1}$ $p_{U_2,X_2}$ $p_{X_3}p_{X_4}$
$p_{Y_3|X_1,X_2,X_4}$ $p_{Y_4|X_1,X_2,X_3}$ $p_{V_3|Y_3}$ $p_{V_4|Y_4}$
(where $p_{Y_3|X_1,X_2,X_4}$ and $p_{Y_4|X_1,X_2,X_4}$ are defined by the
channel) the rate pair $(R_1,R_2)$ is achievable if there are non-negative
$r_{U_1},r_{U_2},r_{X_1},r_{X_2},r_3,r_4$ such that
$R_1=r_{U_1}+r_{X_1}$, $R_2=r_{U_2}+r_{X_2}$, and
\begin{align*}
r_{X_1}&\leq I(X_1;Y_3|X_4,U_1,U_2),\\
r_{U_2}&\leq I(U_2;Y_3,V_4|X_3,X_4,U_1),\\
r_{U_2}&\leq I(U_2;Y_3|X_3,X_4,U_1)
		+ (r_4 - I(Y_4;V_4|X_3,X_4,U_1,U_2,Y_3)),\\
r_{U_1}&\leq I(U_1;Y_3,V_4|X_3,X_4,U_2),\\
r_{U_1}&\leq I(U_1;Y_3|X_3,X_4,U_2)
		+ (r_4 - I(Y_4;V_4|X_3,X_4,U_1,U_2,Y_3)),\\
r_4 &\leq I(X_4;Y_3|U_1,U_2),\\
r_{U_2} + r_{U_1} &\leq I(U_2,U_1;Y_3,V_4|X_3,X_4),\\
r_{U_2} + r_{U_1} &\leq I(U_2,U_1;Y_3|X_3,X_4)
		+ (r_4 - I(Y_4;V_4|X_3,X_4,U_1,U_2,Y_3)),\\
r_4 + r_{U_1}  &\leq I(X_4,U_1;Y_3,V_4|X_3,U_2),\\
r_4 + r_{U_1}  &\leq I(X_4,U_1;Y_3|X_3,U_2)
		+ (r_4 - I(Y_4;V_4|X_3,X_4,U_1,U_2,Y_3)),\\
r_{U_2} + r_4  &\leq I(U_2,X_4;Y_3,V_4|X_3,U_1),\\
r_{U_2} + r_4  &\leq I(U_2,X_4;Y_3|X_3,U_1)
		+ (r_4 - I(Y_4;V_4|X_3,X_4,U_1,U_2,Y_3)),\\
r_{U_2} + r_4 + r_{U_1} &\leq I(U_2,X_4,U_1;Y_3,V_4|X_3),\\
r_{U_2} + r_4 + r_{U_1} &\leq I(U_2,X_4,U_1;Y_3|X_3)
		+ (r_4 - I(Y_4;V_4|X_3,X_4,U_1,U_2,Y_3)),
\end{align*}
and the corresponding inequalities with subscripts 1 and 2 exchanged,
and 3 replaced with 4.
\end{thm}

The theorem is proved in Appendix~\ref{app:destcoopgenericschemes}. A rough
interpretation follows. The auxiliary random variables and the information
associated with them are described below:
\begin{itemize} 

\item {\em {\sf Cooperative public} messages}. $U_1$ and $U_2$ carry the
{\sf cooperative public} messages. These are decoded by both destinations
with help from each other as will become evident.

\item {\em {\sf Private} messages}.  The random variables $X_1$ conditioned
on $U_1$ and $X_2$ conditioned on $U_2$ carry the {\sf private} messages.

\item {\em Quantized observations at the destinations}. The destinations
quantize their observations over each block. The test channel for the
quantizer employed by destination~3 is $p_{V_3|Y_3}$. Similarly,
destination~4 quantizes its observation over a block using the test channel
$p_{V_4|Y_4}$.

\item {\em Messages from destinations to each other}. The quantization codebooks
are binned and the bin-index of the quantized codewords are conveyed by the
destinations to each other in the next block. Destination~3 sends the
bin-index of its quantized codeword to destination~4 in the next block
using $X_3$. Similarly destination~4 quantizes its observation over a block
and conveys the bin-index to destination~3 in the next block using $X_4$.
\end{itemize}

The destinations start decoding from the last block and proceed backwards.
For each block, destination~3 recovers (i) the two {\sf cooperative public}
messages for the current block, (ii) its {\sf private} message for the
current block and (iii) the message from destination~4 which conveys the
bin-index of the quantized codeword at destination~4 for the previous
block. In performing this decoding step, besides its observation for the
current block, destination~3 may rely on the bin-index of the quantized
codeword at destination~4 for the current block since this was recovered in
the previous decoding step. This can be done by the destination jointly
decoding the following:  (i) the two {\sf cooperative public} messages for
the current block, (ii) its {\sf private} message for the current block and
(iii) the message from destination~4 which conveys the bin-index of the
quantized codeword at destination~4 for the previous block, and (iv) the
quantized codeword at destination~4 for the current block. Destination~3
makes use of its observation for the current block and the bin-index of the
quantized codeword at destination~4 for the current block. The {\sf
private} message from the interfering source is treated as noise. Decoding
at destination~4 also proceeds similarly.

As explained in detail in Appendices~\ref{app:destcoopLDachievability}
and~\ref{app:destcoopGachievability} where we prove the achievability part
of our main results, this scheme is useful in the regime where the
cooperative link is stronger than both the direct links from the sources to
their respective destinations. In applying this scheme, we choose the power
levels for the {\sf cooperative public} and {\sf private} signals in a
manner similar to~\cite{EtkinTseWang08}. The {\sf private} signals have a
power level which ensures that they appear at or below the noise-level at
the destinations where they act as interference. The quantizer test-channel
we choose for our purposes is as follows: {\em At destination~3, the
quantization-noise level is equal to the power level at which the {\sf
private} signal from source~1 is received at destination~3.} The intuition
is that since destination~4 is not interested in decoding this {\sf
private} signal from source~1 and treats it as noise, quantizing the
observation at destination~3 any finer than this will not help the joint
decoding significantly. In fact, quantizing more finely could result in
inferior performance since the destinations have to code the messages for
each other at higher rates without producing significant benefits and this
potentially results in overall lower rates for the other messages.

The decoding scheme in Theorem~\ref{thm:destcoopgenericschemes} deviates
slightly from the earlier description. This is done primarily to simplify
the evaluation of the achievable sum-rate for the Gaussian case.  Instead
of decoding the two {\sf cooperative public} messages, the {\sf private}
message, the message from the other receiver, and the quantized codeword at
the other destination for the current block, the destinations initially
decode only (i) the two {\sf cooperative public} messages, (ii) the message
from the other destination, and (iii) the quantized codeword at the
other destination for the current block. This is done treating both the
{\sf private} messages as noise. Once the above messages are recovered, (i)
and (ii) are stripped off from the received signal and the {\sf private}
message is decoded from the residual signal. While this decoding scheme, in
general, could lead to an inferior rate-region compared to the one
described earlier, it is sufficient to obtain the achievability of sum-rate
for the linear deterministic and the Gaussian cases (up to a constant gap
for the Gaussian). For completeness, below we state a generic
achievability theorem which implements the joint decoding described
earlier. We sketch a proof in Appendix~\ref{app:destcoopgeneral}.  However,
we do not use this scheme in proving Theorems~\ref{thm:destcoopLD}
and~\ref{thm:destcoopG}.

\begin{thm}\label{thm:destcoopgeneral}
Given joint distributions $p_{U_1,X_1}$ $p_{U_2,X_2}$ $p_{X_3}p_{X_4}$
$p_{Y_3|X_1,X_2,X_4}$ $p_{Y_4|X_1,X_2,X_3}$ $p_{V_3|Y_3}$ $p_{V_4|Y_4}$
(where $p_{Y_3|X_1,X_2,X_4}$ and $p_{Y_4|X_1,X_2,X_4}$ are defined by the
channel) the rate pair $(R_1,R_2)$ is achievable if there are non-negative
$r_{U_1},r_{U_2},r_{X_1},r_{X_2},r_3,r_4$ such that
$R_1=r_{U_1}+r_{X_1}$, $R_2=r_{U_2}+r_{X_2}$, and
\begin{align*}
r_{X_1}&\leq I(X_1;Y_3,V_4|X_3,X_4,U_1,U_2),\\
r_{X_1}&\leq I(X_1;Y_3|X_3,X_4,U_1,U_2)
\notag\\&\qquad\qquad\qquad\qquad
		+ (r_4 - I(Y_4;V_4|X_3,X_4,X_1,U_2,Y_3)+I(X_1;V_4)),\\
r_4 &\leq I(X_4;Y_3|X_3,X_1,U_2),\\
r_{U_2}+r_{X_1}&\leq I(U_2,X_1;Y_3,V_4|X_3,X_4,U_1),\\
r_{U_2}+r_{X_1}&\leq I(U_2,X_1;Y_3|X_3,X_4,U_1)
\notag\\&\qquad\qquad\qquad\qquad
		+ (r_4 - I(Y_4;V_4|X_3,X_4,X_1,U_2,Y_3)+I(U_2,X_1;V_4)),\\
r_{U_1}+r_{X_1}&\leq I(X_1;Y_3,V_4|X_3,X_4,U_2),\\
r_{U_1}+r_{X_1}&\leq I(X_1;Y_3|X_3,X_4,U_2)
\notag\\&\qquad\qquad\qquad\qquad
		+ (r_4 - I(Y_4;V_4|X_3,X_4,X_1,U_2,Y_3)+I(X_1;V_4)),\\
r_4+r_{X_1} &\leq I(X_4,X_1;Y_3,V_4|X_3,U_2),\\
r_4+r_{X_1} &\leq I(X_4,X_1;Y_3|X_3,U_2)
\notag\\&\qquad\qquad\qquad\qquad
		+ (r_4 - I(Y_4;V_4|X_3,X_4,X_1,U_2,Y_3)+I(X_1;V_4)),\\
r_{U_2}+r_4 &\leq I(U_2,X_4;Y_3,V_4|X_3,X_1),\\
r_{U_2}+r_4 &\leq I(U_2,X_4;Y_3|X_3,X_1)
\notag\\&\qquad\qquad\qquad\qquad
		+ (r_4 - I(Y_4;V_4|X_3,X_4,X_1,U_2,Y_3)+I(U_2;V_4)),\\
r_{U_2} + r_{U_1} + r_{X_1} &\leq I(U_2,X_1;Y_3,V_4|X_3,X_4),\\
r_{U_2} + r_{U_1} + r_{X_1} &\leq I(U_2,X_1;Y_3|X_3,X_4)
\notag\\&\qquad\qquad\qquad\qquad
		+ (r_4 - I(Y_4;V_4|X_3,X_4,X_1,U_2,Y_3)+I(X_1,U_2;V_4)),\\
r_4 + r_{U_1} + r_{X_1} &\leq I(X_4,X_1;Y_3,V_4|X_3,U_2),\\
r_4 + r_{U_1} + r_{X_1} &\leq I(X_4,X_1;Y_3|X_3,U_2)
\notag\\&\qquad\qquad\qquad\qquad
		+ (r_4 - I(Y_4;V_4|X_3,X_4,X_1,U_2,Y_3)+I(X_1;V_4)),\\
r_{U_2} + r_4 + r_{X_1} &\leq I(U_2,X_4,X_1;Y_3,V_4|X_3,U_1),\\
r_{U_2} + r_4 + r_{X_1} &\leq I(U_2,X_4,X_1;Y_3|X_3,U_1)
\notag\\&\qquad\qquad\qquad\qquad
		+ (r_4 - I(Y_4;V_4|X_3,X_4,X_1,U_2,Y_3)+I(X_1,U_2;V_4)),\\
r_{U_2} + r_4 + r_{U_1} + r_{X_1} &\leq I(U_2,X_4,X_1;Y_3,V_4|X_3),\\
r_{U_2} + r_4 + r_{U_1} + r_{X_1} &\leq I(U_2,X_4,X_1;Y_3|X_3)
\notag\\&\qquad\qquad\qquad\qquad
		+ (r_4 - I(Y_4;V_4|X_3,X_4,X_1,U_2,Y_3)+I(X_1,U_2;V_4)),
\end{align*}
and the corresponding inequalities with subscripts 1 and 2 exchanged,
and 3 replaced with 4.
\end{thm}

\section{Upperbounds}
The upperbounds of Theorems~\ref{thm:destcoopLD} and~\ref{thm:destcoopG}
are derived in appendix~\ref{app:upperbounds}. The key ideas behind the
upperbounds are as follows:\\
{\em Upperbound 1:} We simulate two dummy channels which are independent
realizations of the original channel. In both channels, the same codebooks as
in the original are used. In the first dummy channel, the message $M_1$ of
sender 1 is replaced by a dummy message random variable $M_1^\prime$ which
is also uniformly distributed over the alphabet ${\mathcal M}$, but is
independent of both $M_1$ and $M_2$. Similarly, in the second dummy channel
the message $M_2$ is replaced by a dummy message $M_2^{\prime\prime}$. A
genie provides destination~3 with $M_2^{\prime\prime}$ and a certain signal
($g_{1,4}^\ast(X_1^T)+g_{3,4}(X_3^T)$) from the second dummy channel, and
destination~4 with the symmetric counterparts from the first dummy channel.
Applying Fano's inequality and conditions implied by the causality
conditions on the sources, we derive upperbounds which imply
\eqref{eq:Gaussu1} and \eqref{eq:LDu1}.\\
{\em Upperbound 2 and 3:} To show upperbound~2, we consider a genie which
provides $g_{2,3}^\ast(X_2^T)$ and $M_1$ to destination 4 and nothing at
all to destination 3. Using Fano's inequality and using the causality
conditions obeyed by the sources, we derive upperbounds which imply
\eqref{eq:Gaussu2}-\eqref{eq:Gaussu3} and
\eqref{eq:LDu2}-\eqref{eq:LDu3}.\\
\noindent{\em Upperbound 4:} This is a simple cut-set upperbound with
nodes 1 and 3 on one side of the cut and nodes 2 and 4 on the other side.\\
\noindent{\em Upperbound 5:} This is also a cut-set bound. The sources are
on one side of the cut and the destinations on the other.

\section{Discussion}

The gap in Theorem~\ref{thm:destcoopG} can be easily improved by
considering more elaborate schemes and further tightening the upperbound.
We mention a couple of ideas to illustrate how this could be achieved.
However, computing the best possible gap appears to be challenging and we
do not pursue it here.

Even with the schemes we presented in the last section, we picked
potentially sub-optimal power allocations for the different messages
involved in order to simplify the calculations. Improvements in the gap
can be achieved in specific instances simply by optimizing over these power
allocations. But still further improvements can be achieved by  considering
other schemes. For instance, consider the case of a channel with a direct
links which are weak compared to the interfering and cooperative links.
Incorporating a form of decode-and-forward strategy can improve
performance. To see this, let us consider this extreme case:
both direct links are absent
\[ g_{1,3}=g_{2,4}=0,\]
the interfering links have the same strength
\[ |g_{1,4}|=|g_{2,3}|=|g_I|, \] 
and the cooperative links are such that the following condition is satisfied
\[ \left(1+|g_C|^2\right)^2 \leq \left(1+|g_I|^2+|g_C|^2\right).\]
Then, we can show that the sum-capacity is achieved by a simple
decode-and-forward scheme. Both destinations
decode the message from their interfering sources treating the signal in the
cooperative-link as noise, and then in the next block they forward this
decoded message to the other source over the cooperative link. Since the
interference is decoded off first, the signal over the cooperative link can
be decoded without any interference. The resulting rate, under the condition
on the channel strengths mentioned above, is
\[ 2\log\left(1+|g_C|^2\right),\]
which is also what upperbound~\eqref{eq:Gaussu4} works out to. However,
no choice of power allocations in Theorems~\ref{thm:destcoopgenericschemes}
or~\ref{thm:destcoopgeneral} can achieve this. This can be easily remedied by
extending those schemes by incorporating a partial-decode-and-forward
component. However, we do not pursue this direction since the gains are at
most a constant and computing such gains to get an improved uniform bound
appears to be involved.

The upperbounds could also be improved. Modifying the correlation of the
Gaussian noise processes in the additional signals we provide to the
destinations can lead to tighter upperbounds~1, 2, and~3. Also, the
correlation between the input signals can be explicitly accounted for
instead of assuming the worst-case correlation at different stages as we
do. Upperbound~5 can be easily improved by choosing the optimal input
covariance matrix.

\appendices

\section{Proof of Theorem~\ref{thm:destcoopgenericschemes}}
\label{app:destcoopgenericschemes}

We present a block-Markov scheme with backwards decoding.  Given
$p_{U_1,X_1}$ $p_{U_2,X_2}$ $p_{X_3}p_{X_4}$ $p_{Y_3|X_1,X_2,X_4}$
$p_{Y_4|X_1,X_2,X_3}$ $p_{V_3|Y_3}$ $p_{V_4|Y_4}$ (where
$p_{Y_3|X_1,X_2,X_4}$ and $p_{Y_4|X_1,X_2,X_4}$ are defined by the
channel), we construct the following blocklength-$T$ codebooks:

\begin{itemize}

\item {$U$ codebooks:} For $k=1,2$, we create $U_k$-codebooks ${\mathcal
C}_{U_k}$ of size $2^{T(r_{U_k}-3\epsilon)}$ respectively, by choosing
elements independently according to $p_{U_k}$. These codewords will be
denoted by $c_{U_k}(m_{U_k})$ where
$m_{U_k}\in\{1,\ldots,2^{T(r_{U_k}-3\epsilon)}\}$.

\item {$X_1$ and $X_2$ codebooks:} For each codeword $c_{U_k}(m_{U_k})$, we
create a $X_k$-codebook ${\mathcal C}_{X_k}(m_{U_k})$ of size
$2^{T(r_{X_k}-3\epsilon)}$ by choosing elements i.i.d. according to
$p_{X_k|U_k}(.|u_k)$ by setting $u_k$ to be the respective element of the
$c_{U_k}(m_{U_k})$ codeword. We denote these codewords by
$c_{X_k}(m_{X_k},m_{U_k})$, where
$m_{X_k}\in\{1,\ldots,2^{T(r_{X_k}-3\epsilon)}\}$.

\item {$X_3$ and $X_4$ codebooks:} For $k=3,4$, we create $X_k$-codebooks
${\mathcal C}_{X_k}$ of size $2^{T(r_{k}-\epsilon)}$ by choosing the
elements i.i.d. according to $p_{X_k}$. These codewords will be denoted by
$c_{X_k}(m_{X_k})$ where $m_{X_k}\in\{1,\ldots,2^{T(r_{k}-\epsilon)}\}$.

\item {$V_k$ codebooks:} For $k=3,4$, we create $V_k$ codebooks ${\mathcal
C}_{V_k}$ of size $2^{T(I(Y_k;V_k)+\epsilon)}$ by choosing the elements
i.i.d. according to the induced marginal distributions $p_{V_k}$. We bin
these codebooks such that the number of bins is $2^{T(r_{k}-\epsilon)}$.
The codewords will be denoted by $c_{V_k}(b_{V_k},i_{V_k})$ where the
bin-indices are denoted by $b_{V_k}\in\{1,\ldots,2^{T(r_{k}-\epsilon)}\}$,
and within each bin, the index of the codewords are denoted by
$i_{V_k}\in\{1,\ldots,2^{T(I(Y_k;V_k)-r_{k}+2\epsilon)}\}$.

\end{itemize}

\noindent {\em Encoding at the sources:} For block-$j$, $j=1,2,\ldots,J-1$, the
encoders at the sources choose the codewords $c_{U_k}(m_{U_k}(j))$, and
$c_{X_k}(m_{X_k}(j),m_{U_k}(j))$. The $X$-codewords are transmitted. For
the last block $J$, we set
$m_{U_1}(J)=m_{X_1}(J)=m_{U_2}(J)=m_{X_2}(J)=1$.\\

\noindent {\em Encoding at the destinations:} At the end of block-$j$,
$j=1,2,\ldots,J-1$, the destination~3 quantizes its ($T$-length) block of
observations $Y_3^T$ using the $V_3$ codebook by finding a codeword
$c_{V_3}(b_{V_3}(j),i_{V_3}(j))$ which is jointly (strongly)
typical\footnote{In the sequel, we denote the set of strongly
$\delta$-typical sequences by ${\mathcal T}_T^\delta$.
}~\cite[Chapter~13]{CoverThomas} with its observation. If no such codeword
exists, we will say that ``encoding failed at block-$j$'' and declare an
error. However, encoding succeeds with high probability since the $V_3$
codebook has rate $I(V_3;Y_3)$~\cite{WynerZiv}. Then destination~3 sets
$m_{X_3}(j+1)=b_{V_3}(j)$ and for block-$j+1$, destination~3 sends
$c_{X_3}(m_{X_3}(j+1))$. The encoding at destination~4 proceeds
similarly.\\

\noindent {\em Decoding at the destinations:} Destinations perform {\em
backwards decoding}~\cite{WillemsBackwards}. We will assume that before
destination~3 processes block-$j$, it has already successfully decoded
$m_{X_4}(j+1)$. This is true with high probability\footnote{{\em i.e.},
with probability approaching~1 as the blocklength~$T$ goes to $\infty$.}
for $j=J$ if
\[ r_4 \leq I(X_4; Y_3|X_1,X_2).\]

For each $j=1,2,\ldots, J-1$, we will ensure that from block-$j$,
destination~3 decodes $m_{X_4}(j)$ successfully with high probability
thereby ensuring that the above assumption holds true. Assuming that
$m_{X_4}(j+1)$, which is equal to $b_{V_4}(j)$, is available at
destination~3, we will ensure that from the observation $Y_3$ made by
destination~3 in block-$j$, the messages $m_{U_1}(j)$, $m_{X_1}(j)$, and
$m_{X_4}(j)$ can be successfully decoded with high probability. The
decoding will proceed in two steps. In the first step, destination~3 will
attempt to decode the messages $m_{U_1}(j)$ and $m_{X_4}(j)$ along with the
message $m_{U_2}(j)$. Then, conditioned on these messages, it will try to
decode the message $m_{X_1}(j)$. Concretely, in the first step the decoder
looks for a unique collection of codewords such that they are jointly
typical with its observation $Y_3^T(j)$ and the information it already has,
namely $m_{X_3}(j)$ and $b_{V_4}(j)$. In other words, destination~3
searches for a unique $(\hat{m}_{U_1}(j), \hat{m}_{U_2}(j),
\hat{m}_{X_4}(j))$ such that
\[ ({c}_{U_1}(\hat{m}_{U_1}(j)),{c}_{U_2}(\hat{m}_{U_2}(j)),
 {c}_{X_4}(\hat{m}_{X_4}(j)),{c}_{X_3}(m_{X_3}(j)),{c}_{V_4}(b_{V_4}(j),
 \hat{i}_{V_4}(j)),Y_3^T(j)) \in {\mathcal T}_T^\delta,\]
for some $\hat{i}_{V_4}(j)$. We will argue below that this decoding
succeeds, {\em i.e.}, $(\hat{m}_{U_1}(j), \hat{m}_{U_2}(j),
\hat{m}_{X_4}(j)) = ({m}_{U_1}(j),{m}_{U_2}(j),{m}_{X_4}(j))$ with high
probability if the following conditions are met.
\begin{align*}
r_{U_2}&\leq I(U_2;Y_3,V_4|X_3,X_4,U_1),\\
r_{U_2}&\leq I(U_2;Y_3|X_3,X_4,U_1)
		+ (r_4 - I(Y_4;V_4|X_3,X_4,U_1,U_2,Y_3)),\\
r_{U_1}&\leq I(U_1;Y_3,V_4|X_3,X_4,U_2),\\
r_{U_1}&\leq I(U_1;Y_3|X_3,X_4,U_2)
		+ (r_4 - I(Y_4;V_4|X_3,X_4,U_1,U_2,Y_3)),\\
r_4 &\leq I(X_4;Y_3|U_1,U_2),\\
r_{U_2} + r_{U_1} &\leq I(U_2,U_1;Y_3,V_4|X_3,X_4),\\
r_{U_2} + r_{U_1} &\leq I(U_2,U_1;Y_3|X_3,X_4)
		+ (r_4 - I(Y_4;V_4|X_3,X_4,U_1,U_2,Y_3)),\\
r_4 + r_{U_1}  &\leq I(X_4,U_1;Y_3,V_4|X_3,U_2),\\
r_4 + r_{U_1}  &\leq I(X_4,U_1;Y_3|X_3,U_2)
		+ (r_4 - I(Y_4;V_4|X_3,X_4,U_1,U_2,Y_3)),\\
r_{U_2} + r_4  &\leq I(U_2,X_4;Y_3,V_4|X_3,U_1),\\
r_{U_2} + r_4  &\leq I(U_2,X_4;Y_3|X_3,U_1)
		+ (r_4 - I(Y_4;V_4|X_3,X_4,U_1,U_2,Y_3)),\\
r_{U_2} + r_4 + r_{U_1} &\leq I(U_2,X_4,U_1;Y_3,V_4|X_3),\\
r_{U_2} + r_4 + r_{U_1} &\leq I(U_2,X_4,U_1;Y_3|X_3)
		+ (r_4 - I(Y_4;V_4|X_3,X_4,U_1,U_2,Y_3)).
\end{align*}
In the second step, destination~3 decodes $m_{X_1}(j)$ using its
observation $Y_3^T(j)$ and what it decoded in the previous step, namely,
$\hat{m}_{U_1}(j)$, $\hat{m}_{U_2}(j)$, and $\hat{m}_{X_4}(j)$. It looks
for a unique $\hat{m}_{X_1}(j)$ such that
\[ (c_{X_1}(\hat{m}_{X_1}(j)),{c}_{U_1}(\hat{m}_{U_1}(j)),
    c_{U_2}(\hat{m}_{U_2}(j)),{c}_{X_4}(\hat{m}_{X_4}(j)),Y_3^T(j)) \in 
    {\mathcal T}_T^\delta.\]
Assuming that the first step succeeded, it can be shown that the second
step succeeds in decoding the correct message with a high probability if
\[ r_{X_1} \leq I(X_1;Y_3|U_1,U_2,X_4).\]

We will now argue that the probability of error in the first step is
vanishingly small. We first note that the correct choice of messages will
result in a jointly typical set of codewords with high probability. {\em
i.e.}, when $T\rightarrow \infty$,
\[{\mathbb P}\bigg(
\left({c}_{U_1}({m}_{U_1}(j)),{c}_{U_2}({m}_{U_2}(j)),
 {c}_{X_4}({m}_{X_4}(j)),{c}_{X_3}(m_{X_3}(j)),{c}_{V_4}(b_{V_4}(j),
 {i}_{V_4}(j)),Y_3^T(j)\right) \in {\mathcal T}_T^\delta\bigg) \rightarrow 1.\]
This is essentially a statement of Markov Lemma~\cite[Lemma~4.1]{Berger77}
(also see~\cite[Corollary~3.2.3.1]{TungPhD}). We need to show that the
probability of the event ${\sf E}$ that there is some $(\hat{m}_{U_1}(j),
\hat{m}_{U_2}(j), \hat{m}_{X_4}(j)) \neq ({m}_{U_1}(j), {m}_{U_2}(j),
{m}_{X_4}(j))$ and some $\hat{i}_{V_4}$ such that
\[ \left({c}_{U_1}(\hat{m}_{U_1}(j)),{c}_{U_2}(\hat{m}_{U_2}(j)),
 {c}_{X_4}(\hat{m}_{X_4}(j)),{c}_{X_3}(m_{X_3}(j)),{c}_{V_4}(b_{V_4}(j),
 \hat{i}_{V_4}(j)),Y_3^T(j)\right) \in {\mathcal T}_T^\delta\]
is small. This event is the union of the following two events: (1) where
$\hat{i}_{V_4}$ takes on its correct value $i_{V_4}$, and (2) where
$\hat{i}_{V_4}\neq i_{V_4}$. Further, each of these events are unions of
events where some or none (but not all) of the messages take on their
correct value. We apply union bound to upperbound ${\mathbb P}({\sf E})$.
To illustrate, let us consider two events:
\begin{align*}
{\sf E}_{1,2,4}&=\left\{ \hat{m}_{U_1}(j)\neq {m}_{U_1}(j),
\hat{m}_{U_2}(j)\neq {m}_{U_2}(j), \hat{m}_{X_4}(j) \neq
{m}_{X_4}(j),\hat{i}_{V_4}(j) = i_{V_4}(j) \right\},\\
{\sf E}^\ast_{1,2,4}&=\left\{ \hat{m}_{U_1}(j)\neq {m}_{U_1}(j),
\hat{m}_{U_2}(j)\neq {m}_{U_2}(j), \hat{m}_{X_4}(j) \neq
{m}_{X_4}(j),\hat{i}_{V_4}(j) \neq i_{V_4}(j) \right\}.
\end{align*}
We have
\begin{align*}
&{\mathbb P}({\sf E}_{1,2,4})\\
&=\sum_{\small \begin{array}{c}\hat{m}_{U_1}\neq {m}_{U_1}(j),\\ 
 \hat{m}_{U_2}\neq {m}_{U_2}(j),\\ \hat{m}_{X_4} \neq {m}_{X_4}(j)
 \end{array}} {\mathbb P}\left( 
 ({c}_{U_1}(\hat{m}_{U_1}), {c}_{U_2}(\hat{m}_{U_2}), 
 {c}_{X_4}(\hat{m}_{X_4}), {c}_{X_3}(m_{X_3}(j)), {c}_{V_4}(b_{V_4}(j),
 {i}_{V_4}(j)),Y_3^T(j)) \in {\mathcal T}_T^\delta \right)\\
&\leq 2^{T(r_{U_1}+r_{U_2}+r_{4}-7\epsilon)}
2^{T(-I(U_1,U_2,X_4;X_3,V_4,Y_3)+\delta)}\\
&= 2^{T(r_{U_1}+r_{U_2}+r_{4}-I(U_1,U_2,X_4;V_4,Y_3|X_3) -7\epsilon +
\delta)}.
\end{align*}
If the rates satisfy the condition
\[ r_{U_1}+r_{U_2}+r_{4}\leq I(U_1,U_2,X_4;V_4,Y_3|X_3),\]
the probability of the error event ${\sf E}_{1,2,4}$ can be made
vanishingly small. Similarly,
\begin{align*}
&{\mathbb P}({\sf E}^\ast_{1,2,4})\\
&=
 \sum_{\small \begin{array}{c} \hat{m}_{U_1}\neq {m}_{U_1}(j),\\
 \hat{m}_{U_2}\neq {m}_{U_2}(j),\\ \hat{m}_{X_4} \neq {m}_{X_4}(j),\\
 \hat{i}_{V_4}\neq i_{V_4}(j)\end{array}} {\mathbb P}\left( 
 ({c}_{U_1}(\hat{m}_{U_1}), {c}_{U_2}(\hat{m}_{U_2}), 
 {c}_{X_4}(\hat{m}_{X_4}), {c}_{X_3}(m_{X_3}(j)), {c}_{V_4}(b_{V_4}(j),
 \hat{i}_{V_4}),Y_3^T(j)) \in {\mathcal T}_T^\delta \right)\\
 &= 2^{T(r_{U_1}+r_{U_2}+r_4+(I(Y_4;V_4)-r_4)-5\epsilon)}
    2^{T(-I(U_1,U_2,X_4,V_4;X_3,Y_3)+\delta)}\\
 &= 2^{T((r_{U_1}+r_{U_2}+r_4 - I(U_1,U_2,X_4;X_3,Y_3)) -
         (r_4-I(V_4;Y_4)+I(V_4;X_3,Y_3|U_1,U_2,X_4)) -5\epsilon+\delta)}
\end{align*}
To drive the probability of this error event (${\sf E}^\ast_{1,2,4}$) to
zero, it is enough to ensure that
\begin{align*}
r_{U_1} +r_{U_2} + r_4 \leq I(U_1,U_2,X_4;X_3,Y_3) +
(r_4-I(V_4;Y_4)+I(V_4;X_3,Y_3|U_1,U_2,X_4)).
\end{align*}
Note that
\begin{align*}
I(V_4;Y_4)-I(V_4;X_3,Y_3|U_1,U_2,X_4)) &=
  I(V_4;Y_4,U_1,U_2,X_4,X_3,Y_3)-I(V_4;X_3,Y_3|U_1,U_2,X_4))\\
 &= I(V_4;U_1,U_2,X_4) + I(V_4;Y_4|X_3,X_4,U_1,U_2,Y_3)\\
 &\geq I(V_4;Y_4|X_3,X_4,U_1,U_2,Y_3),
\end{align*}
where the first equality follows from the fact that $V_4 - Y_4 - (U_1, U_2,
X_4, X_3, Y_3)$ is a Markov chain. Hence, ${\mathbb P}({\sf
E}^\ast_{1,2,4})$ can be made small if
\[ r_{U_1}+r_{U_2}+r_4 \leq I(U_1,U_2,X_4;Y_3|X_3) + (r_4 -
I(V_4;Y_4|X_3,X_4,U_1,U_2,Y_3)).\]
Similarly, considering the other possible error events results in the rest
of the conditions. A similar set of conditions ensure success of decoding
at destination~4. If decoding fails for block-$j$ for either of the
destinations, we will say that ``decoding failed at block-$j$'' and declare
an error.

Overall, an error results if for at least one block-$j$, either encoding
fails or decoding fails. Since there are a finite number $J$ of blocks, by
union bound, the above discussion implies that the probability of error
goes to 0 as the blocklength goes to $\infty$ when the above conditions are
met. This completes the random coding argument.

\section{Proof sketch of Theorem~\ref{thm:destcoopgeneral}}
\label{app:destcoopgeneral}

The codebook construction and encoding at the sources and destinations are
identical to the one in Appendix~\ref{app:destcoopgenericschemes}. The only
difference is in how the messages are decoded by the destinations. The
destinations again follow a backwards decoding procedure similar to the one
there. However, instead of carrying it out in two steps, the destinations
attempt to decode the same set of codewords as there, but in a single step.
In particular, destination~3 while decoding block-$j$ looks for a unique
set of  $(\hat{m}_{U_1}(j), \hat{m}_{U_2}(j), \hat{m}_{X_4}(j),
\hat{m}_{X_1}(j))$ such that
\[ ({c}_{U_1}(\hat{m}_{U_1}(j)),{c}_{U_2}(\hat{m}_{U_2}(j)),
 {c}_{X_4}(\hat{m}_{X_4}(j)),{c}_{X_1}(\hat{m}_{U_1}(j)),
 {c}_{X_3}(m_{X_3}(j)),{c}_{V_4}(b_{V_4}(j),
 \hat{i}_{V_4}(j)),Y_3^T(j)) \in {\mathcal T}_T^\delta,\]
for some $\hat{i}_{V_4}(j)$. Note that, as in
Appendix~\ref{app:destcoopgenericschemes}, in performing this decoding
step, destinations~3 makes uses of the bin-index $b_{V_4}(j)$ which was
recovered from processing block-$j+1$. The conditions on the rates in
Theorem~\ref{thm:destcoopgeneral} ensure that the probability of all
relevant error events are small for sufficiently large values of $T$. The
analysis is along the same lines as in
Appendix~\ref{app:destcoopgenericschemes} and is omitted.

\section{Proof of the achievability of Theorem~\ref{thm:destcoopLD}}
\label{app:destcoopLDachievability}

If we fix $n_{1,3}$, $n_{1,4}$, $n_{2,3}$, and $n_{2,4}$, and consider the
$u_i$'s in \eqref{eq:LDu1}-\eqref{eq:LDu4} as functions of $n_C$, the
sum-rate expression in Theorem~\ref{thm:destcoopLD} (as a function of
$n_C$) breaks up into three natural regimes. We use different strategies to
achieve the sum-capacity in different regimes. The regimes are:
\begin{enumerate}
\item[(i)] $n_C\leq
n_\text{min}\defineqq\min(n_{1,3},n_{1,4},n_{2,3},n_{2,4})$.
It can be shown that for $n_C\geq n_\text{min}$,
\[u_1(n_C) \geq \min(u_2(n_C), u_3(n_C), u_4(n_C), u_5).\]
Hence, we need consider $u_1$ only in the regime $n_C\leq
n_\text{min}$. Moreover, in this regime, $u_2(n_C)$ through $u_4(n_C)$ are
constants ({\em i.e.}, they do not depend on $n_C$ and their values are the
same as when $n_C=0$). Since $u_1(n_C)$ is monotonically increasing in
$n_C$, this means that we need to employ cooperation only when $u_1(0) <
\min(u_2(0),u_3(0),u_4(0),u_5)$, {\em i.e.}, when
\begin{align}
\max&(n_{1,3}-n_{1,4},n_{2,3})+\max(n_{2,4}-n_{1,4},n_{1,4})\notag\\ &\quad<
\min\bigg(\max(n_{1,3},n_{2,3}) + \left(\max(n_{2,4},n_{2,3})-n_{2,3}\right),
\notag\\ &\qquad\qquad\qquad
\max(n_{2,4},n_{1,4}) + \left(\max(n_{1,3},n_{1,4})-n_{1,4}\right), 
n_{1,3}+n_{2,4}\bigg).\label{eq:LDu1cond}
\end{align}
When the above condition is not true, the sum-rate expression reduces to
the sum-capacity without cooperation. 

\item[(ii)] $n_\text{min}< n_C \leq \min(n_{1,3},n_{2,4})$. In this regime,
we can observe that the sum-rate expression takes on a constant value since
$u_2(n_C)$, $u_3(n_C)$, and $u_4(n_C)$ are still constants. Hence, the
achievability here is implied by the achievability in regime (i).

\item[(iii)] $\min(n_{1,3},n_{2,4})< n_C$. In
this regime, we use Theorem~\ref{thm:destcoopgenericschemes}.

\end{enumerate}
For integer $q$ satisfying $1\leq q\leq n$, we define
\begin{align*}
{\mathcal F}_{q}\stackrel{\text{def}}{=}\left\{ x\in
{\mathbb F}^n: x_i=0,\; i\leq q\right\},
\end{align*}
{\em i.e.}, all vectors in ${\mathbb F}^n$ such that their components
in the range $1,\ldots,q$ are zeros. We take the indexing of the
elements of vectors to start from the top as usual. For example, for binary field
and $n=4$,
\[{\mathcal F}_2=\left\{
\left[\begin{array}{c}0\\0\\0\\0\end{array}\right],
\left[\begin{array}{c}0\\0\\0\\1\end{array}\right],
\left[\begin{array}{c}0\\0\\1\\0\end{array}\right],
\left[\begin{array}{c}0\\0\\1\\1\end{array}\right]\right\}.\]

\noindent{\em Regime~(i): $n_C \leq n_{i,j}$, $i\in\{1,2\}, j\in\{3,4\}.$}\\
First of all, we note that if 
\[ n_{1,4}+n_{2,3} \geq n_{1,3}+n_C \text{ and }
 n_{1,4}+n_{2,3} \geq n_{2,4}+n_C, \]
then, 
\[ u_1(n_C) = u_1(0).\]
Putting this together with the fact that $u_2(n_C), u_3(n_C),$ and
$u_4(n_C)$ are independent of $n_C$ in regime~(i) (as discussed above), we
may conclude that it is enough to show achievability under no cooperative
link ({\em i.e.}, under $n_C=0$). But this achievability is already known
(see~\cite{SourceCoop}, for instance).
Hence, we will assume that at least one of the following two conditions is
true.
\[ n_{1,4}+n_{2,3} > n_{1,3}+n_C \text{ and }
 n_{1,4}+n_{2,3} > n_{2,4}+n_C. \]
We will first consider case (a) where the following conditions are
satisfied.
\[ n_{1,4}+n_{2,3} < n_{1,3}+n_C \text{ and } n_{1,4}+n_{2,3} < n_{2,4}+n_C. \]
Following that, we will consider case~(b) where
\[ n_{1,4}+n_{2,3} < n_{1,3}+n_C \text{ and } 
   n_{1,4}+n_{2,3} \geq n_{2,4}+n_C. \]
By symmetry, this would also cover the third possibility of 
\[ n_{1,4}+n_{2,3} \geq n_{1,3}+n_C \text{ and } 
   n_{1,4}+n_{2,3} < n_{2,4}+n_C. \]

For case (a), let us consider the following block-Markov scheme with
superposition coding. Let $U_k,S_k,Z_{\up k},Z_{\down k},\;k=1,2$ be
independent auxiliary random variables with marginal distributions
$p_{U_k},p_{S_k},p_{Z_{\up k}},p_{Z_{\down k}}$. The alphabet for these
random variables is ${\mathbb F}^n$. Corresponding to these random
variables, random codebooks of blocklength-$T$ and rates $r_{U_k}, r_{S_k},
r_{Z_{\up k}}, r_{Z_{\down k}}$, respectively, are defined as usual. For
instance, the $U_1$-codebook, denoted by ${\mathcal C}_{U_1}$, is of size
$2^{T(r_{U_1}-\epsilon)}$ is generated by choosing the $T$ elements of each
of the codewords independently according to $p_{U_1}$.  These codewords
will be denoted by $c_{U_1}(m_{U_1})$ where
$m_{U_1}\in\{1,\ldots,2^{T(r_{U_1}-\epsilon)}\}$. The blocks will be
indexed by $j=1,2,\ldots,J$. The message transmitted by source~1 using the
$U_1$-codebook in block-$j$ will be denoted by $m_{U_1}(j)$, and the
corresponding codeword by $c^{(j)}_{U_1}=c_{U_1}(m_{U_1}(j))$.
For block-$j$, sources transmit the following blocklength-$T$ vectors
\begin{align*}
X^{(j)}_1&=c^{(j)}_{U_1} + c^{(j)}_{Z_{\up 1}} + c^{(j)}_{Z_{\down 1}} +
f^{(j)}_{S_1},\\
X^{(j)}_2&=c^{(j)}_{U_2} + c^{(j)}_{Z_{\up 2}} + c^{(j)}_{Z_{\down 2}} + 
f^{(j)}_{S_2},
\end{align*}
where
\begin{align*}
f^{(j)}_{S_1} = c^{(j)}_{S_1} + {\bf A}_1 c^{(j+1)}_{S_1} + \ldots + {\bf A}_1^{J-j}
c^{(J)}_{S_1},\\
f^{(j)}_{S_2} = c^{(j)}_{S_2} + {\bf A}_2 c^{(j+1)}_{S_2} + \ldots + {\bf A}_2^{J-j}
c^{(J)}_{S_2},
\end{align*}
where ${\bf A}_1$ and ${\bf A}_2$ are matrices given below. The addition is vector
addition.
\begin{align*}
{\bf A}_1&={\bf S}^{n_{1,3}+n_C-n_{1,4}-n_{2,3}},\quad&\quad
{\bf A}_2&={\bf S}^{n_{2,4}+n_C-n_{1,4}-n_{2,3}}.
\end{align*}
In the sequel we will ensure that the rates of the codebooks are such that
from observing $Y_3^{(j)}$, the $j$-th block observed by destination~3, it
(destination~3) can decode with a high probability of success the codewords
$c^{(j)}_{U_1}$, $c^{(j)}_{S_1}$, and $c^{(j)}_{Z_{\up 1}}$, for all $j$.
Similarly, we will make sure that destination~4 will successfully decode
$c^{(j)}_{U_2}$, $c^{(j)}_{S_2}$, and $c^{(j)}_{Z_{\up 2}}$ from
$Y_4^{(j)}$. Then, the destinations will transmit, respectively, for
$j=1,2,\ldots,J-1$,
\begin{align*}
X_3^{(j+1)}={\bf A}_3\left(Y_3^{(j)}
 -{\bf G}_{1,3}\left(c^{(j)}_{U_1} + c^{(j)}_{S_1}+c^{(j)}_{U_{\up 1}}\right)
 -{\bf G}_C{\bf A}_4{\bf G}_{1,4}\left(c^{(j-1)}_{U_1}+c^{(j-1)}_{Z_{\up 1}}\right) \right),\\
X_4^{(j+1)}={\bf A}_4\left(Y_4^{(j)}
  -{\bf G}_{2,4}\left(c^{(j)}_{U_2} + c^{(j)}_{S_2}+c^{(j)}_{U_{\up 2}}\right)
  -{\bf G}_C{\bf A}_3{\bf G}_{2,3}\left(c^{(j-1)}_{U_2}+c^{(j-1)}_{Z_{\up 2}}\right)\right),
\end{align*}
where, ${\bf G}_{k_1,k_2}$ is a short-hand notation for ${\bf
S}^{n-n_{k_1,k_2}}$, and ${\bf G}_C$ for ${\bf S}^{n-n_C}$. Also, ${\bf A}_3$ and ${\bf A}_4$
are matrices defined below.
\begin{align*}
{\bf A}_4&=-{\bf S}^{-(n-n_{1,4})},\\
{\bf A}_3&=-{\bf S}^{-(n-n_{2,3})}.
\end{align*}
The choices for the distributions of the auxiliary random variables used to
create the codebooks will be taken up in the sequel.

With these, the received signals at the destinations are
\begin{align*}
Y_3^{(j)}&=\wbar{g_{1,3}}^{(j)}(c_{U_1} + c_{Z_{\up 1}}) 
    + {\bf G}_{1,3}(f^{(j)}_{S_1})
    + \wtilde{g_{1,3}}^{(j)}(c_{Z_{\down 1}}) 
    + {\bf G}_{2,3}(c^{(j)}_{U_2}+c^{(j)}_{Z_{\up 2}})
    + \wtilde{g_{2,3}}^{(j)}(c_{Z_{\down 2}}),\\
Y_4^{(j)}&=\wbar{g_{2,4}}^{(j)}(c_{U_2} + c_{Z_{\up 2}}) 
    + {\bf G}_{2,4}(f^{(j)}_{S_2})
    + \wtilde{g_{2,4}}^{(j)}(c_{Z_{\down 2}}) 
    + {\bf G}_{1,4}(c^{(j)}_{U_1}+c^{(j)}_{Z_{\up 1}})
    + \wtilde{g_{1,4}}^{(j)}(c_{Z_{\down 1}}),
\end{align*}
where the functions are as defined below. Note that $Y_3$ does not have any
terms which depend on $S_2$-codewords (and similarly, $Y_4$ does not
involve any terms containing $S_1$-codewords). This was achieved by the
appropriate choices above for ${\bf A}_1$ and ${\bf A}_4$ (respectively, ${\bf A}_2$ and ${\bf A}_3$).
\begin{align*}
\wbar{g_{1,3}}^{(j)}(c_{U_1}+c_{Z_{\up 1}}) &=
  {\bf G}_{1,3}(c^{(j)}_{U_1}+c^{(j)}_{Z_{\up 1}}) 
  + {\bf G}_C{\bf A}_4{\bf G}_{1,4}(c^{(j-1)}_{U_1}+c^{(j-1)}_{Z_{\up 1}}),\\
\wtilde{g_{1,3}}^{(j)}(c_{Z_{\down 1}}) &=
  \sum_{i\in\{j,j-2,\ldots,1\}}({\bf G}_C{\bf A}_4{\bf G}_C{\bf A}_3)^{\frac{j-i}{2}}
                                 {\bf G}_{1,3}c^{(i)}_{Z_{\down 1}}\\
&\qquad\qquad\qquad\qquad
  + \sum_{i\in\{j-1,j-3,\ldots,1\}} ({\bf G}_C{\bf A}_4{\bf G}_C{\bf A}_3)^{\frac{j-1-i}{2}}
                                    {\bf G}_C {\bf A}_4{\bf G}_{1,4}c^{(i)}_{Z_{\down 1}},\\
\wtilde{g_{2,3}}^{(j)}(c_{Z_{\down 2}}) &=
  \sum_{i\in\{j,j-2,\ldots,1\}} ({\bf G}_C{\bf A}_4{\bf G}_C{\bf A}_3)^{\frac{j-i}{2}}
                                {\bf G}_{2,3}c^{(i)}_{Z_{\down 2}}\\
&\qquad\qquad\qquad\qquad
  + \sum_{i\in\{j-1,j-3,\ldots,1\}} ({\bf G}_C{\bf A}_4{\bf G}_C{\bf A}_3)^{\frac{j-1-i}{2}}
                                   {\bf G}_C {\bf A}_4  {\bf G}_{2,4} c^{(i)}_{Z_{\down 2}},\\
\end{align*}
and the functions $\wbar{g_{2,4}}^{(j)}(c_{U_2}+c_{Z_{\up 1}})$,
$\wtilde{g_{2,4}}^{(j)}(c_{Z_{\down 2}})$, and 
$\wtilde{g_{1,4}}^{(j)}(c_{Z_{\down 1}})$ are defined similarly.
Note that, as before, ${\bf G}_{k_1,k_2}$ is a short-hand notation for
${\bf S}^{n-n_{k_1,k_2}}$, and ${\bf G}_C$ for ${\bf S}^{n-n_C}$.

As mentioned earlier, destinations perform decoding in two phases. At the
end of every block $j$, the destinations decode the $c^{(j)}_{U}$,
$c^{(j)}_{S}$ and $c^{(j)}_{Z_{\up}}$ intended for them (in that order). We
call this {\em phase~1 decoding}. At the end of block-$J$, the decoders
perform a {\em phase~2 decoding} where it decodes the following codewords
for all blocks, {\em i.e.}, for each $j=1,2,\ldots, J$, the destinations
decode  $c^{(j)}_{U}$ intended for the other user and $c^{(j)}_{Z_{\down}}$
intended for itself (in that order).  In both of the phases, the decodings
are performed in the order mentioned above treating all the undecoded
codewords and other interference as noise. Below, we will specify the
distributions employed and evaluate the conditions on the rates to ensure
successful decoding. This will establish achievability for regime~(i).

The auxiliary random variables $U_1, S_1, Z_{\up 1},$ and $Z_{\down 1}$,
respectively, are uniformly distributed over ${\mathbb F}^n$,
${\mathcal F}_{n_{1,4}-n_C}$, ${\mathcal F}_{n_{1,4}}$, and ${\mathcal
F}_{n_{1,3}+n_C-n_{2,3}}$.
Destination~3 transmits
\begin{align*}
X^{(j)}_3 &= {\bf A}_3\left(Y_3^{(j)}-{\bf G}_{1,3}\left(c^{(j)}_{U_1} + 
c^{(j)}_{S_1}+c^{(j)}_{U_{\up 1}}\right)\right)\\
 &= {\bf A}_3\left( \wtilde{g_{1,3}}^{(j)}(c_{Z_{\down 1}}) 
    + {\bf G}_{2,3}(c^{(j)}_{U_2}+c^{(j)}_{Z_{\up 2}})
    + \wtilde{g_{2,3}}^{(j)}(c_{Z_{\down 2}}) \right)
\end{align*}
The rates supported by the above scheme for source~1 are given by the
following set of conditions.
\begin{align*}
r_{U_1} &\leq [n_{1,4}-n_C],\\
r_{U_1} &\leq n_C,\\
r_{S_1} &\leq n_C - [n_{2,3}-(n_{1,3}-n_{1,4})]_+,\\
r_{Z_{\up 1}} &\leq [n_{1,3}-n_{1,4}-n_{2,3}]_+,\\
r_{Z_{\down 1}} &\leq [n_{2,3}-n_C]_+.
\end{align*}
The first constraint on $r_{U_1}$ comes from the phase~1 decoding at
destination~3, while the second condition is from the phase~2 decoding at
destination~4. A similar set of constraints apply for the rates achievable
by source~2.

Combining all these, an achievable sum-rate is given by
\begin{align*}
R_\text{sum} &= \max(n_{1,3}-n_{1,4}+n_C,n_{1,3}-n_C) +
\max(n_{2,4}-n_{2,3}+n_C,n_{2,4}-n_C).
\end{align*}
This, combined with the fact that the achievability of a given sum-rate at
a lower value of $n_C$ implies its achievability for all larger values of
$n_C$ provided the rest of the channel coefficients remain the same
allows us to conclude that the minimum of the following three terms is
achievable.
\begin{align*}
&n_{1,3}-n_{1,4}+n_C + n_{2,4}-n_{2,3}+n_C,\\
&n_{1,3} + n_{2,4} - n_{1,4}, \text{ and}\\
&n_{2,4} + n_{1,3} - n_{2,3}.
\end{align*}
Under case~(a), it is easy to verify that this is precisely what the
upperbound evaluates to. Hence, we have shown achievability for case~(a).

Let us now consider case (b) where 
\[ n_{1,4}+n_{2,3} < n_{1,3}+n_C \text{ and } 
   n_{1,4}+n_{2,3} \geq n_{2,4}+n_C. \]
The achievable strategy we use for this case involves source~1 transmitting
according to a scheme similar to the one above while source~2 employs a
superposition coding scheme similar to that of Han and Kobayashi for the
interference channel without a cooperative link. In particular, only node~3
uses its transmission capabilities.

The codebooks and the choice of distributions for
source~1 are exactly as above, except for the choice of rates which will be
presented in the sequel. Source~2 uses only the following codebooks:
$U_2$ and $Z_{\down 2}$. Moreover, $Z_{\down 2}$ is now uniformly
distributed over ${\mathcal F}_{n_{2,3}}$.
Exactly as in the earlier scheme, destination~3 performs a two-phase decoding
and transmits a shifted version of the residual signals after the first
phase of decoding of the previous block. The shift matrix ${\bf A}_3$ is the
same as above. Destination~4, on the other hand, performs only a single
phase of decoding where $U_2$, $U_1$ and $Z_{\down 2}$ codewords are
decoded. As mentioned earlier, destination~4 does not transmit anything,
{\em i.e.}, ${\bf A}_4=0$. The received signals can be seen to be
\begin{align*}
Y_3^{(j)}&={\bf G}_{1,3}(c^{(j)}_{U_1} + c^{(j)}_{Z_{\up 1}} + f^{(j)}_{S_1}
 + c^{(j)}_{Z_{\down 1}}) + {\bf G}_{2,3}(c^{(j)}_{U_2}+c^{(j)}_{Z_{\down 2}}),\\
Y_4^{(j)}&={\bf G}_{2,4}(c^{(j)}_{U_2} + c^{(j)}_{Z_{\down 2}})
    + {\bf G}_C {\bf A}_3 {\bf G}_{2,3} (c^{(j-1)}_{U_2} + c^{(j-1)}_{Z_{\down 2}})
    + {\bf G}_{1,4}(c^{(j)}_{U_1}+c^{(j)}_{Z_{\up 1}})
    + \wtilde{g_{1,4}}^{(j)}(c_{Z_{\down 1}}),
\end{align*}
where, unlike earlier,
\begin{align*}
\wtilde{g_{1,4}}^{(j)}(c_{Z_{\down 1}}) &= {\bf G}_{1,4}c^{(j)}_{Z_{\down 1}}
  + {\bf G}_C {\bf A}_3  {\bf G}_{1,3} c^{(j-1)}_{Z_{\down 1}}.
\end{align*}
Note that, again the appropriate choice of ${\bf A}_1$ and ${\bf A}_3$ has
ensured that no contribution from the $S_1$ codeword is observed at
destination~4.

Decoding at destination~3 proceeds as in the above scheme. On the other
hand, destination~4 decodes, at the end of each block, (i) first, the $U_2$
and $U_1$ codewords jointly treating all other signals and interference as
noise, and then from the residual signal (ii) $Z_{\down 2}$ codeword
treating interference as noise. The conditions on the rates for successful
decoding are given below.
\begin{align*}
r_{U_1} &\leq [n_{1,4}-n_C],\\
r_{U_2} &\leq n_C,\\
r_{S_1} &\leq n_C - [n_{2,3}-(n_{1,3}-n_{1,4})]_+,\\
r_{Z_{\up 1}} &\leq [n_{1,3}-n_{1,4}-n_{2,3}]_+,\\
r_{Z_{\down 1}} &\leq [n_{2,3}-n_C]_+,\\
r_{U_2} &\leq  \min(n_{2,3},n_{2,4}),\\
r_{U_1} &\leq [n_{1,4}-[n_{2,4}-n_{2,3}]_+],\\
r_{U_2}+r_{U_1} &\leq \max(n_{2,4},n_{1,4})-[n_{2,4}-n_{2,3}]_+,\\
r_{Z_{\down 2}} &\leq [n_{2,4}-n_{2,3}]_+.
\end{align*}
where the first five conditions ensure successful decoding at destination~3
and the rest of the three conditions does the same for decoding at
destination~4.
Upon simplifying, we may conclude that a sum-rate equal to the minimum of
the following terms is achievable
\begin{align*}
&n_{1,3} + [n_{2,4}-n_{2,3}]_+,\\
&n_{1,3} - n_{1,4} + \max(n_{1,4},n_{2,4}), \text{and }\\
&n_{1,3} + n_C.
\end{align*}
It is easy to check that this is what our upperbound evaluates to under
case~(b). Thus, we have also proved achievability under case~(b).

\noindent{\em Regime (iii): $\min(n_{1,3},n_{2,4})< n_C$.} We employ
Theorem~\ref{thm:destcoopgenericschemes} with the following choices for
$p_{U_1,X_1}$, $p_{U_2,X_2}$, $p_{X_3}$, $p_{X_4}$, $p_{V_3|Y_3}$, and
$p_{V_4|Y_4}$: $U_1,U_2,X_3,X_4$ are independent and identically
distributed uniformly over ${\mathbb F}^n$. $Z_1$ and $Z_2$, respectively,
are uniformly distributed over ${\mathcal F}_{n_{1,4}}$ and ${\mathcal
F}_{n_{2,3}}$, respectively. They are independent of $U_1,U_2,X_3,X_4$ and
of each other. We define
\begin{align*}
 X_1 &= U_1 + Z_1,\\
 X_2 &= U_2 + Z_2.
\end{align*}
$p_{V_3|Y_3}$ and $p_{V_4|Y_4}$ are defined by the following deterministic 
{\em test-channels}
\begin{align*}
 V_3 &= {\bf S}^{[n_{1,3}-n_{1,4}]_+} Y_3,\\
 V_4 &= {\bf S}^{[n_{2,4}-n_{2,3}]_+} Y_4.
\end{align*}
Note that this amounts to the destinations truncating their observations to
the level at which their own {\sf private}-codewords ({\em e.g.}, $Z_1$ in
the case of destination~3) are received. Thus, the quantized observations
$V$'s only contain information on the {\sf public}-codewords (the $U$'s).
This is consistent with the intuition that the information they forward on
to the other destination is utilized to recover the {\sf public}-codewords.

With these choices, the conditions on the non-negative rates for
achievability from Theorem~\ref{thm:destcoopgenericschemes} work out to
\begin{align*}
r_{X_1}&\leq [n_{1,3}-n_{1,4}]_+,\\
r_{U_1}&\leq \max\left( n_{1,3} - [n_{1,3}-n_{1,4}]_+, 
                        n_{1,4} - [n_{2,4}-n_{2,3}]_+ \right),\\
r_{U_1}&\leq n_{1,3} - [n_{1,3}-n_{1,4}]_+ + r_4,\\
r_{U_2}&\leq \max\left( n_{2,3} - [n_{1,3}-n_{1,4}]_+, 
                        n_{2,4} - [n_{2,4}-n_{2,3}]_+ \right),\\
r_{U_2}&\leq  [n_{2,3} - [n_{1,3}-n_{1,4}]_+]_+ + r_4,\\
r_4 &\leq [n_C - [n_{1,3}-n_{1,4}]_+]_+,\\
r_{U_1} + r_{U_2} &\leq \left\{ \begin{array}{l}
\max\bigg( (n_{1,3} - [n_{1,3}-n_{1,4}]_+) +
           (n_{2,4} - [n_{2,4}-n_{2,3}]_+),\\
\qquad\qquad 
           [n_{1,4} - [n_{2,4}-n_{2,3}]_+]_+ +
           [n_{2,3} - [n_{1,3}-n_{1,4}]_+]_+ \bigg),\\
 \qquad\qquad\qquad\qquad\qquad\qquad\qquad\qquad
 \text{if }n_{1,3}+n_{2,4} \neq n_{1,4}+n_{2,3}\\
\max\bigg( (n_{1,3} - [n_{1,3}-n_{1,4}]_+),
           (n_{2,4} - [n_{2,4}-n_{2,3}]_+),\\
\qquad\qquad 
           (n_{1,4} - [n_{2,4}-n_{2,3}]_+),
           (n_{2,3} - [n_{1,3}-n_{1,4}]_+) \bigg),\\
 \qquad\qquad\qquad\qquad\qquad\qquad\qquad\qquad
 \text{otherwise,}
\end{array}\right.\\
r_{U_1} + r_{U_2} &\leq \max( n_{1,3},n_{2,3} ) -
                        [n_{1,3}-n_{1,4}]_+ + r_4,\\
r_{U_1} + r_4  &\leq \max\bigg( \max(n_{1,3},n_C) - [n_{1,3}-n_{1,4}]_+,\\
 &\qquad\qquad\qquad\qquad\qquad\qquad
 [n_{1,4}-[n_{2,4}-n_{2,3}]_+]_+ + [n_C - [n_{1,3}-n_{1,4}]_+]_+\bigg),\\
r_{U_1} + r_4  &\leq n_{1,3} - [n_{1,3}-n_{1,4}]_+ + r_4,\\
r_{U_2} + r_4  &\leq \max\bigg( \max(n_{2,3},n_C) - [n_{1,3}-n_{1,4}]_+,\\
 &\qquad\qquad\qquad\qquad\qquad\qquad
 (n_{2,4}-[n_{2,4}-n_{2,3}]_+) + [n_C - [n_{1,3}-n_{1,4}]_+]_+\bigg),\\
r_{U_2} + r_4  &\leq [n_{2,3} - [n_{1,3}-n_{1,4}]_+]_+ + r_4,
\end{align*}
\begin{align*}
r_{U_2} + r_4 + r_{U_1} &\leq \left\{ \begin{array}{l}
\max\bigg( (n_{1,3} - [n_{1,3}-n_{1,4}]_+) +
           (n_{2,4} - [n_{2,4}-n_{2,3}]_+),\\
 \qquad\qquad
           [n_{1,4} - [n_{2,4}-n_{2,3}]_+]_+ +
           [n_{2,3} - [n_{1,3}-n_{1,4}]_+]_+,\\
 \qquad\qquad
           [n_C - [n_{1,3}-n_{1,4}]_+]_+ + 
           (\max(n_{1,4},n_{2,4}) - [n_{2,4}-n_{2,3}]_+) \bigg),\\
 \qquad\qquad\qquad\qquad\qquad\qquad\qquad\qquad\qquad\quad
 \text{if }n_{1,3}+n_{2,4} \neq n_{1,4}+n_{2,3}\\
\max\bigg( (n_{1,3} - [n_{1,3}-n_{1,4}]_+),\;
           (n_{2,4} - [n_{2,4}-n_{2,3}]_+),\\
 \qquad\qquad
           (n_{1,4} - [n_{2,4}-n_{2,3}]_+),
           (n_{2,3} - [n_{1,3}-n_{1,4}]_+),\\
 \qquad\qquad
           [n_C - [n_{1,3}-n_{1,4}]_+]_+ + 
           (\max(n_{1,4},n_{2,4}) - [n_{2,4}-n_{2,3}]_+) \bigg),\\
 \qquad\qquad\qquad\qquad\qquad\qquad\qquad\qquad\qquad\quad
 \text{otherwise,}
\end{array}\right. \\
r_{U_2} + r_4 + r_{U_1} &\leq \max( n_{1,3},n_{2,3},n_C ) - 
                               [n_{1,3}-n_{1,4}]_+ + r_4,
\end{align*}
and the corresponding inequalities with subscripts 1 and 2 exchanged,
and 3 replaced with 4. Applying Fourier-Motzkin elimination to these
conditions, we can show that a sum-rate equal to the minimum of the
following terms is achievable in this regime.
\begin{align*}
u_2&=
\max(n_{2,4},n_{2,3}) + \left(\max(n_{1,3},n_{2,3},n_C)-n_{2,3}\right),\\
u_3&=
\max(n_{1,3},n_{1,4}) + \left(\max(n_{2,4},n_{1,4},n_C)-n_{1,4}\right),\\
u_4&=\max(n_{1,3},n_C)+\max(n_{2,4},n_C),\text{ and}\\
u_5&=\left\{\begin{array}{ll}
\max(n_{1,3}+n_{2,4},n_{1,4}+n_{2,3}), 
 &\text{ if }n_{1,3}-n_{2,3}\neq n_{1,4}-n_{2,4},\\
\max(n_{1,3},n_{2,4},n_{1,4},n_{2,3}),
 &\text{ otherwise}.
\end{array}\right.
\end{align*}

\section{Proof of the achievability of Theorem \ref{thm:destcoopG}}
\label{app:destcoopGachievability}

The proof of Theorem~\ref{thm:destcoopG} will follow the proof of
Theorem~\ref{thm:destcoopLD} closely. We first make the following
definitions:
\begin{align*}
n_{k_1,k_2}&\defineqq [\log|g_{k_1,k_2}|^2]_+,\;k_1,k_2\in\{ 1,2,3,4\},\text{ and}\\
n_C &\defineqq [\log|g_C|^2]_+.
\end{align*}
Let us observe that the minimum of the following four terms $u'_1, u'_2,
u'_3$, and $u'_4$ are within a constant (7 bits) of the minimum of
the corresponding unprimed terms, $u_1,u_2,u_3$, and $u_4$.
\begin{align*}
u'_1&=\max(n_{1,3}-n_{1,4}+n_C,n_{2,3},n_C) +
\max(n_{2,4}-n_{2,3}+n_C,n_{1,4},n_C),\\
u'_2&=
\max(n_{1,3},n_{2,3}) + \left(\max(n_{2,4},n_{2,3},n_C)-n_{2,3}\right),\\
u'_3&=
\max(n_{2,4},n_{1,4}) + \left(\max(n_{1,3},n_{1,4},n_C)-n_{1,4}\right),\\
u'_4&=\max(n_{1,3},n_C)+\max(n_{2,4},n_C).
\end{align*} 
Hence, it is enough to show that the minimum of the four terms above and 
\begin{align*} u'_5 =
\log\bigg( 1 &+ 
 \left(|g_{1,3}|^2+|g_{2,4}|^2+|g_{1,4}|^2+|g_{2,3}|^2\right)\\ 
&\qquad\qquad+
 \left( |g_{1,3}g_{2,4}|^2 + |g_{1,4}g_{2,3}|^2 -
2|g_{1,3}g_{2,4}g_{1,4}g_{2,3}|\cos\theta\right) \bigg), 
\end{align*}
which is also within a constant (2 bits) of $u_5$, is achievable. We again
consider the same three regimes as in
Appendix~\ref{app:destcoopLDachievability}.

\noindent{\em Regime~(i): $|g_C| \leq |g_{i,j}|$, $i\in\{1,2\}, j\in\{3,4\}.$}\\

We will assume that $|g_C|>1$. If $|g_C|\leq 1$, it can be verified that
our upperbound is not more than 4 bits away from the upperbound for the
corresponding Gaussian interference channel without a cooperation link
({\em i.e.} $g_C=0$) in~\cite{EtkinTseWang08} which itself is known to be
achievable with a gap of at most 2 bits. Hence overall, if $|g_C|\leq 1$,
the upperbound is achievable with a gap of 6 bits.

Further, we will show achievability only for the case where $2|g_C| \leq
\min(|g_{1,4}|,|g_{2,3}|)$. Note that the upperbounds change by at most 2
bits if we do not impose this restriction (but still maintain the
restriction that $|g_C| \leq |g_{i,j}|,\; i\in\{1,2\}, j\in\{3,4\}$), and
achievability of a given sum-rate at a lower value of $|g_C|$ implies its
achievability for all larger values of $|g_C|$ ({\em i.e.}, the
sum-capacity is monotonic in $|g_C|$) provided the rest of the channel
coefficients remain the same. Hence, showing achievability under this
restricted $|g_C|$ regime implies a proof of achievability for the whole
regime with a further gap of 2 bits from the upperbound.

Let us define
\begin{align*}
A_1&\defineqq\left|\frac{g_{1,4}g_{2,3}}{g_{1,3}g_C}\right|,\\
A_2&\defineqq\left|\frac{g_{1,4}g_{2,3}}{g_{2,4}g_C}\right|.
\end{align*}
We first note that in regime (i) with the additional assumptions we made
above, if both $A_1, A_2 \geq 1/2$, then our
upperbound can be shown to be not more than 10 bits from the upperbound
in~\cite{EtkinTseWang08} for the Gaussian interference channel without a
cooperative link. Since that upperbound is known to be achievable within
two bits, the gap to the upperbound is at most 12. Hence, we will only
consider the other three possibilities: (a) $A_1< 1/2,\quad A_2< 1/2$, (b)
$A_1< 1/2,\quad A_2\geq 1/2$, and (c) $A_1\geq 1/2,\quad A_2< 1/2$. We will
show achievability for cases~(a) and (b). Case~(c) will follow from
case~(b) by symmetry.

The coding scheme will very closely resemble the one we used for regime~(i)
in Appendix~\ref{app:destcoopLDachievability}. We repeat all the details
below for completeness.

For case~(a), let us consider the following block-Markov scheme with
superposition coding. Let $U_k,S_k,Z_{\up k},Z_{\down k},\;k=1,2$ be
independent auxiliary random variables with marginal distributions
$p_{U_k},p_{S_k},p_{Z_{\up k}},p_{Z_{\down k}}$. The alphabet for these
random variables is the set of complex numbers. Corresponding to these
random variables, random codebooks of blocklength-$T$ and rates $r_{U_k},
r_{S_k}, r_{Z_{\up k}}, r_{Z_{\down k}}$, respectively, are defined as
usual. For instance, the $U_1$-codebook, denoted by ${\mathcal C}_{U_1}$,
is of size $2^{T(r_{U_1}-\epsilon)}$ is generated by choosing the $T$
elements of each of the codewords independently according to $p_{U_1}$.
These codewords will be denoted by $c_{U_1}(m_{U_1})$ where
$m_{U_1}\in\{1,\ldots,2^{T(r_{U_1}-\epsilon)}\}$. The blocks will be
indexed by $j=1,2,\ldots,J$. The message transmitted by source~1 using the
$U_1$-codebook in block-$j$ will be denoted by $m_{U_1}(j)$, and the
corresponding codeword by $c^{(j)}_{U_1}=c_{U_1}(m_{U_1}(j))$.

For block-$j$, sources transmit the following blocklength-$T$ vectors
\begin{align*}
X^{(j)}_1&=c^{(j)}_{U_1} + c^{(j)}_{Z_{\up 1}} + c^{(j)}_{Z_{\down 1}} +
f^{(j)}_{S_1},\\
X^{(j)}_2&=c^{(j)}_{U_2} + c^{(j)}_{Z_{\up 2}} + c^{(j)}_{Z_{\down 2}} + 
f^{(j)}_{S_2},
\end{align*}
where
\begin{align*}
f^{(j)}_{S_1} = c^{(j)}_{S_1} + A_1 c^{(j+1)}_{S_1} + \ldots + A_1^{J-j}
c^{(J)}_{S_1},\\
f^{(j)}_{S_2} = c^{(j)}_{S_2} + A_2 c^{(j+1)}_{S_2} + \ldots + A_2^{J-j}
c^{(J)}_{S_2}.
\end{align*}

In the sequel we will ensure that the rates of the codebooks are such that
from observing $Y_3^{(j)}$, the $j$-th block observed by destination~3, it
(destination~3) can decode with a high probability of success the codewords
$c^{(j)}_{U_1}$, $c^{(j)}_{S_1}$, and $c^{(j)}_{Z_{\up 1}}$, for all $j$.
Similarly, we will make sure that destination~4 will successfully decode
$c^{(j)}_{U_2}$, $c^{(j)}_{S_2}$, and $c^{(j)}_{Z_{\up 2}}$ from
$Y_4^{(j)}$. Then, the destinations will transmit, respectively, for
$j=1,2,\ldots,J-1$,
\begin{align*}
X_3^{(j+1)}=A_3\left(Y_3^{(j)}
 -g_{1,3}\left(c^{(j)}_{U_1} + c^{(j)}_{S_1}+c^{(j)}_{U_{\up 1}}\right)
 -g_CA_4g_{1,4}\left(c^{(j-1)}_{U_1}+c^{(j-1)}_{Z_{\up 1}}\right) \right),\\
X_4^{(j+1)}=A_4\left(Y_4^{(j)}
  -g_{2,4}\left(c^{(j)}_{U_2} + c^{(j)}_{S_2}+c^{(j)}_{U_{\up 2}}\right)
  -g_CA_3g_{2,3}\left(c^{(j-1)}_{U_2}+c^{(j-1)}_{Z_{\up 2}}\right)\right),
\end{align*}
where,  $A_3$ and $A_4$ are defined below
\begin{align*}
A_4&= -\frac{g_{2,3}}{g_C g_{2,4} A_2},\\
A_3&=-\frac{g_{1,4}}{g_C g_{1,3} A_1},\\
\end{align*}
The choices for the distributions and the fact that
the power constraints are satisfied will be taken up in the sequel.

With these, the received signals at the destinations are
\begin{align*}
Y_3^{(j)}&=\wbar{g_{1,3}}^{(j)}(c_{U_1} + c_{Z_{\up 1}}) 
    + g_{1,3}(f^{(j)}_{S_1})
    + \wtilde{g_{1,3}}^{(j)}(c_{Z_{\down 1}}) 
    + g_{2,3}(c^{(j)}_{U_2}+c^{(j)}_{Z_{\up 2}})
    + \wtilde{g^\ast_{2,3}}^{(j)}(c_{Z_{\down 2}}),\\
Y_4^{(j)}&=\wbar{g_{2,4}}^{(j)}(c_{U_2} + c_{Z_{\up 2}}) 
    + g_{2,4}(f^{(j)}_{S_2})
    + \wtilde{g_{2,4}}^{(j)}(c_{Z_{\down 2}}) 
    + g_{1,4}(c^{(j)}_{U_1}+c^{(j)}_{Z_{\up 1}})
    + \wtilde{g^\ast_{1,4}}^{(j)}(c_{Z_{\down 1}}),
\end{align*}
where the functions are as defined below. Note that $Y_3$ does not have any
terms which depend on $S_2$-codewords (and similarly, $Y_4$ does not
involve any terms containing $S_1$-codewords). This was achieved by the
appropriate choices above for $A_1$ and $A_4$ (respectively, $A_2$ and $A_3$).
Here
\begin{align*}
\wbar{g_{1,3}}^{(j)}(c_{U_1}+c_{Z_{\up 1}}) &=
  g_{1,3}(c^{(j)}_{U_1}+c^{(j)}_{Z_{\up 1}}) 
  + g_CA_4g_{1,4}(c^{(j-1)}_{U_1}+c^{(j-1)}_{Z_{\up 1}}),\\
g_{2,3}(c^{(j)}_{U_2}+c^{(j)}_{Z_{\up 2}}) &=
  g_{2,3}(c^{(j)}_{U_2}+c^{(j)}_{Z_{\up 2}}),
\end{align*}
\begin{align*}
\wtilde{g_{1,3}}^{(j)}(c_{Z_{\down 1}}) &=
  \sum_{i\in\{j,j-2,\ldots,1\}}(g_CA_4g_CA_3)^{\frac{j-i}{2}}
                                 g_{1,3}c^{(i)}_{Z_{\down 1}}
  + \sum_{i\in\{j-1,j-3,\ldots,1\}} (g_CA_4g_CA_3)^{\frac{j-1-i}{2}}
                                    g_C A_4g_{1,4}c^{(i)}_{Z_{\down 1}},\\
\wtilde{g^\ast_{2,3}}^{(j)}(c_{Z_{\down 2}}) &=
  \sum_{i\in\{j,j-2,\ldots,1\}} (g_CA_4g_CA_3)^{\frac{j-i}{2}}
                                g_{2,3}c^{(i)}_{Z_{\down 2}}
  + \sum_{i\in\{j-1,j-3,\ldots,1\}} (g_CA_4g_CA_3)^{\frac{j-1-i}{2}}
                                   g_C A_4  g_{2,4} c^{(i)}_{Z_{\down 2}}\\
  &\quad+ \sum_{i\in\{j,j-2,\ldots,1\}} (g_CA_4g_CA_3)^{\frac{j-i}{2}}
                                  N^{(i)}_{1}
  + \sum_{i\in\{j-1,j-3,\ldots,1\}} g_C A_4 (g_CA_4g_CA_3)^{\frac{j-1-i}{2}}
                                    N^{(i)}_{2},
\end{align*}
and the functions $\wtilde{g_{2,4}}^{(j)}(c_{Z_{\down 2}})$, 
$g_{1,4}(c^{(j)}_{U_1}+c^{(j)}_{Z_{\up 1}})$, and 
$\wtilde{g^\ast_{1,4}}^{(j)}(c_{Z_{\down 1}})$ are defined similarly.

As before, destinations perform decoding in two phases. At the
end of every block $j$, the destinations decode the $c^{(j)}_{U}$,
$c^{(j)}_{S}$ and $c^{(j)}_{Z_{\up}}$ intended for them (in that order). We
call this {\em phase~1 decoding}. At the end of block-$J$, the decoders
perform a {\em phase~2 decoding} where it decodes the following codewords
for all blocks, {\em i.e.}, for each $j=1,2,\ldots, J$, the destinations
decode  $c^{(j)}_{U}$ intended for the other user and $c^{(j)}_{Z_{\down}}$
intended for itself (in that order).  In both of the phases, the decodings
are performed in the order mentioned above treating all the undecoded
codewords and other interference as noise. Below, we will specify the
distributions employed and evaluate the conditions on the rates to ensure
successful decoding. This will establish achievability for regime~(i).

The auxiliary random variables are all Gaussian with
the following powers: 
\begin{align*}
\sigma_{U_1}^2&=1/K,\\
\sigma_{S_1}^2&=\frac{1}{K}\left|\frac{g_C}{g_{1,4}}\right|^2,\\
\sigma_{Z_{\up 1}}^2&= \frac{1}{K}\left|\frac{1}{g_{1,4}}\right|^2,\\
\sigma_{Z_{\down 1}}^2 &= 
   \frac{1}{K}\left|\frac{g_{2,3}}{g_{1,3}g_C}\right|^2,
\end{align*}
where $K$ is a constant which will be chosen presently to satisfy the power
constraint. The power allocation for the auxiliary random variables for
source~2 is chosen similarly.

Destination~3 transmits
\begin{align*}
X^{(j)}_3 &= A_3\left(Y_3^{(j)}-g_{1,3}\left(c^{(j)}_{U_1} + 
c^{(j)}_{S_1}+c^{(j)}_{U_{\up 1}}\right)\right)\\
 &= A_3\left( \wtilde{g_{1,3}}^{(j)}(c_{Z_{\down 1}}) 
    + g_{2,3}(c^{(j)}_{U_2}+c^{(j)}_{Z_{\up 2}})
    + \wtilde{g^\ast_{2,3}}^{(j)}(c_{Z_{\down 2}}) \right)
\end{align*}
If $2|g_C| \leq |g_{1,4}|, |g_{2,3}|$ and $|g_C|\geq 1$, under the above
power allocation, the average power of these terms can be
shown to be
\begin{align*}
(1/T)\|A_3\wtilde{g_{1,3}}^{(j)}(c_{Z_{\down 1}})\|^2 
    &\leq \frac{|g_{2,3}|^2(2/K)}{|g_C|^2}\\
(1/T)\|A_3g_{2,3}(c^{(j)}_{U_2}+c^{(j)}_{Z_{\up 2}})\|^2 
    &\leq 2/K\\
(1/T){\mathbb E}\left[\|A_3\wtilde{g^\ast_{2,3}}^{(j)}(c_{Z_{\down
2}})\|^2\right] 
    &\leq \frac{(2/K) + 2}{|g_{2,3}|^2}.
\end{align*}
We can easily verify from this that if we choose $K=9$, the power
constraints at all the transmitters are satisfied. The rates supported by
the above scheme for source~1 are given by the following set of conditions
\begin{align*}
r_{U_1}&\leq \log\left( 1 + \frac{\frac{|g_{1,3}|^2}{K}}
           {\frac{2|g_{1,3}g_C|^2}{K|g_{1,4}|^2} +
\frac{|g_{1,3}|^2}{K|g_{1,4}|^2} +
            \frac{2|g_{2,3}|^2}{K|g_C|^2} + \frac{2|g_{2,3}|^2}{K} +
\frac{2}{K} + 2} \right),\\
r_{U_1}&\leq \log\left( 1 + \frac{\frac{|g_{1,4}|^2}{K}}
         {\frac{|g_{1,4}|^2}{K|g_C|^2} + \frac{2}{K} + \frac{2}{K} + 2}\right),\\
r_{S1}&\leq  \log\left( 1 + \frac{\frac{|g_{1,3}g_C|^2}{K|g_{1,4}|^2}}
    {\frac{2|g_{2,3}|^2}{K} + \frac{|g_{1,3}|^2}{K|g_{1,4}|^2} +
\frac{2|g_{2,3}|^2}{K|g_C|^2} + \frac{2|g_{2,3}|^2}{K} + \frac{2}{K} + 2} \right),\\
r_{Z_{\up 1}}&\leq \log\left( 1 + \frac{\frac{|g_{1,3}|^2}{K|g_{1,4}|^2}}
     {\frac{2|g_{2,3}|^2}{K} + \frac{2|g_{2,3}|^2}{K|g_C|^2} +
\frac{2|g_{2,3}|^2}{K} + \frac{2}{K} + 2} \right),\\
r_{Z_{\down 1}}&\leq \log\left( 1 + \frac{\frac{|g_{2,3}|^2}{K|g_C|^2}}
    {\frac{2}{K} + 2}\right).
\end{align*}
The first constraint on $r_{U_1}$ comes from the phase~1 decoding at
destination~3, while the second condition is from the phase~2 decoding at
destination~4. A similar set of constraints apply for the rates achievable
by source~2.

Simplifying, these constraints imply that rates which satisfy the following
are also achievable
\begin{align*}
r_{U_1} &\leq [n_{1,4}-n_C] - 4,\\
r_{U_1} &\leq n_C - 4,\\
r_{S_1} &\leq n_C - [n_{2,3}-(n_{1,3}-n_{1,4})]_+ - 4,\\
r_{Z_{\up 1}} &\leq [n_{1,3}-n_{1,4}-n_{2,3}]_+ - 4,\\
r_{Z_{\down 1}} &\leq [n_{2,3}-n_C]_+ - 5.
\end{align*}

Combining all these, an achievable sum-rate is given by
\begin{align*}
R_\text{sum} &= \max(n_{1,3}-n_{1,4}+n_C,n_{1,3}-n_C) +
\max(n_{2,4}-n_{2,3}+n_C,n_{2,4}-n_C) - 34
\end{align*}
This, combined with the fact that the achievability of a given sum-rate at
a lower value of $|g_C|$ implies its achievability for all larger values of
$|g_C|$ provided the rest of the channel coefficients remain the same
allows us to conclude that the minimum of the following three terms is
achievable with a gap of 34.
\begin{align*}
&n_{1,3}-n_{1,4}+n_C + n_{2,4}-n_{2,3}+n_C,\\
&n_{1,3} + n_{2,4} - n_{1,4}, \text{ and}\\
&n_{2,4} + n_{1,3} - n_{2,3}.
\end{align*}
The above minimum is also the minimum of $u'_1,u'_2,u'_3$ under case~(a).
This shows achievability in case~(a).

Let us now consider case (b) where $A_1<1/2$, but $A_2\geq 1/2$. The
achievable strategy we use for this case involves source~1 transmitting
according to a scheme similar to the one above while source~2 employs a
superposition coding scheme similar to that of Han and Kobayashi for the
interference channel without a cooperative link. In particular, only node~3
uses its transmission capabilities.

The codebooks and the choice of distributions (power allocations) for
source~1 is exactly as above, except for the choice of rates which will be
presented in the sequel.  Source~2 uses only the following codebooks:
$U_2$ and $Z_{\down 2}$. Moreover, the choice of $\sigma_{Z_{\down 2}}^2$ is
different.
\[ \sigma_{Z_{\down 2}}^2 = \frac{1}{K}\frac{1}{|g_{2,3}|^2}.\]
Exactly as in the above scheme, destination~3 performs a two-phase decoding
and transmits a scaled version of the residual signals after the first
phase of decoding of the previous block. The scaling factor $A_3$ is the
same as above. Destination~4, on the other hand, performs only a single
phase of decoding where $U_2$, $U_1$ and $Z_{\down 2}$ codewords are
decoded. As mentioned earlier, destination~4 does not transmit anything,
{\em i.e.}, $A_4=0$. The received signals can be seen to be
\begin{align*}
Y_3^{(j)}&=g_{1,3}(c^{(j)}_{U_1} + c^{(j)}_{Z_{\up 1}} + f^{(j)}_{S_1}
 + c^{(j)}_{Z_{\down 1}}) + g_{2,3}^\ast(c^{(j)}_{U_2}+c^{(j)}_{Z_{\down 2}}),\\
Y_4^{(j)}&=g_{2,4}(c^{(j)}_{U_2} + c^{(j)}_{Z_{\down 2}})
    + g_C A_3 g_{2,3} (c^{(j-1)}_{U_2} + c^{(j-1)}_{Z_{\down 2}})
    + g_{1,4}(c^{(j)}_{U_1}+c^{(j)}_{Z_{\up 1}})
    + \wtilde{g^\ast_{1,4}}^{(j)}(c_{Z_{\down 1}}),
\end{align*}
where, unlike above,
\begin{align*}
\wtilde{g^\ast_{1,4}}^{(j)}(c_{Z_{\down 1}}) &= g_{1,4}c^{(j)}_{Z_{\down 1}}
  + g_C A_3  g_{1,3} c^{(j-1)}_{Z_{\down 1}}
  + N^{(j)}_{2} +  g_C A_3 N^{(j-1)}_{2},
\end{align*}
and the other functions are as before. Note that, again the appropriate
choice of $A_1$ and $A_3$ has ensured that no contribution from the $S_1$
codeword is observed at destination~4. The fact that the power constraints
are satisfied at all the transmitters under the earlier choice of $K=9$ is
easy to verify.

Decoding at destination~3 proceeds as in the above scheme. Whereas,
destination~4 decodes, at the end of each block, (i) first, the $U_2$ and
$U_1$ codewords jointly  treating all other signals and interference as
noise, and then from the residual signal (ii) $Z_{\down 2}$ codeword
treating interference as noise. The conditions on the rates for successful
decoding are given below.
\begin{align*}
r_{U_1}&\leq \log\left( 1 + \frac{|g_{1,3}|^2/K}
           {2|g_{1,3}g_C|^2/(K|g_{1,4}|^2) + |g_{1,3}|^2/(K|g_{1,4}|^2) +
            |g_{2,3}|^2/(K|g_C|^2) + |g_{2,3}|^2 + 1} \right),\\
r_{S_1}&\leq  \log\left( 1 + \frac{|g_{1,3}g_C|^2/(K|g_{1,4}|^2)}
    {2|g_{2,3}|^2/K + |g_{1,3}|^2/(K|g_{1,4}|^2) + |g_{2,3}|^2/(K|g_C|^2)
    + |g_{2,3}|^2 + 1} \right),\\
r_{Z_{\up 1}}&\leq \log\left( 1 + \frac{|g_{1,3}|^2/(K|g_{1,4}|^2)}
     {2|g_{2,3}|^2/K + |g_{2,3}|^2/(K|g_C|^2) + |g_{2,3}|^2 
      + 1} \right),\\
r_{U_2}&\leq \log\left( 1 + \frac{|g_{2,3}|^2/K}
      {|g_{2,3}|^2/(K|g_C|^2) + 1 + 1}\right),\\
r_{Z_{\down 1}}&\leq \log\left( 1 + \frac{|g_{2,3}|^2/(K|g_C|^2)}
    {1 + 1}\right),\\
\\
r_{U_2}&\leq \log\left( 1 + \frac{|g_{2,4}|^2/K}
{|g_{2,4}|^2/(K|g_{2,3}|^2) + 1/K + 3/2K + 3/2}\right),\\
r_{U_1}&\leq \log\left( 1 + \frac{|g_{1,4}|^2/K}
         {|g_{2,4}|^2/(K|g_{2,3}|^2) + 1/K + 3/2K + 3/2}\right),\\
r_{U_2}+r_{U_1} &\leq \log\left( 1 + \frac{|g_{2,4}|^2/K+|g_{1,4}|^2/K}
         {|g_{2,4}|^2/(K|g_{2,3}|^2) + 1/K + 3/2K + 3/2}\right),\\
r_{Z_{\down 2}} &\leq \log\left( 1 + \frac{|g_{2,4}|^2/(K|g_{2,3}|^2)}
         {1/K + 3/2K + 3/2}\right),
\end{align*}
where the first five conditions ensure successful decoding at destination~3
and the rest of the three conditions does the same for decoding at
destination~4. We may simplify the terms to conclude that rates which
satisfy all the conditions below are achievable.
\begin{align*}
r_{U_1} &\leq [n_{1,4}-n_C] - 4,\\
r_{U_2} &\leq n_C - 4,\\
r_{S_1} &\leq n_C - [n_{2,3}-(n_{1,3}-n_{1,4})]_+ - 4,\\
r_{Z_{\up 1}} &\leq [n_{1,3}-n_{1,4}-n_{2,3}]_+ - 4,\\
r_{Z_{\down 1}} &\leq [n_{2,3}-n_C]_+ - 5,\\
r_{U_2} &\leq  \min(n_{2,3},n_{2,4}) - 5,\\
r_{U_1} &\leq [n_{1,4}-[n_{2,4}-n_{2,3}]_+] - 5,\\
r_{U_2}+r_{U_1} &\leq \max(n_{2,4},n_{1,4})-[n_{2,4}-n_{2,3}]_+ - 5,\\
r_{Z_{\down 2}} &\leq [n_{2,4}-n_{2,3}]_+ - 4.
\end{align*}
Upon simplifying, we may conclude that a sum-rate equal to the minimum of
the following terms is achievable within 28 bits
\begin{align*}
&n_{1,3} + [n_{2,4}-n_{2,3}]_+,\\
&n_{1,3} - n_{1,4} + \max(n_{1,4},n_{2,4}), \text{and }\\
&n_{1,3} + n_C.
\end{align*}
This is the minimum of $u'_1, u'_2,$ and $u'_3$ under case~(b). Thus, we
have shown achievability under case~(b) as well.

Overall, we have shown achievability of the upperbound in regime~(i) with a
gap of at most 43 bits.

\noindent{\em Regime (ii):} As in
Appendix~\ref{app:destcoopLDachievability}, achievability in regime~(i)
implies the achievability in regime~(ii) as well since in this regime
\begin{align*}
u'_1(n_C) \geq \min(u'_2(n_C),u'_3(n_C),u'_4(n_C),u'_5),
\end{align*}
and $u'_2(n_C)$, $u'_3(n_C)$, $u'_4(n_C)$ and $u'_5$ are constants.

\noindent{\em Regime (iii):} 
Note that we proved Theorem~\ref{thm:destcoopgenericschemes} for discrete
alphabets, but the extension to the continuous alphabet case is standard
and we will assume that version for proving achievability here.

We apply Theorem~\ref{thm:destcoopgenericschemes} with the following
choices for the auxiliary random variables. $Z_1,Z_2,U_1,U_2,X_3,X_4$ are
independent and identically distributed zero-mean Gaussian random
variables. The variances are, respectively
\begin{align*}
\sigma_{Z_1}^2&=\frac{1/K}{\max(1,|g_{1,4}|^2)},\\
\sigma_{Z_2}^2&=\frac{1/K}{\max(1,|g_{2,3}|^2)},\\
\sigma_{U_1}^2=\sigma_{U_2}^2&={1/K},\\
\sigma_{X_3}^2=\sigma_{X_4}^2&=1,\\
\end{align*}
where we set $K=1/2$. Further,
\begin{align*}
 X_1 &= U_1 + Z_1,\\
 X_2 &= U_2 + Z_2.
\end{align*}
It is easy to see that this satisfies the power constraint since $K=1/2$.
$p_{V_3|Y_3}$ and $p_{V_4|Y_4}$ are defined by the following 
{\em test-channels}
\begin{align*}
 V_3 &= Y_3 + Q_3,\text{ and}\\
 V_4 &= Y_4 + Q_4,
\end{align*}
where $Q_3$ and $Q_4$ are independent, zero-mean Gaussian random variables
which are also independent of $Y_3,Y_4$ and all the other auxiliary random
variables. Their variances are, respectively
\begin{align*}
\sigma_{Q_3}^2 &=
  \max\left(1, \frac{\max(1,|g_{1,3}|^2)}{\max(1,|g_{1,4}|^2)}\right),
   \text{ and}\\
\sigma_{Q_4}^2 &=
  \max\left(1, \frac{\max(1,|g_{2,4}|^2)}{\max(1,|g_{2,3}|^2)}\right).
\end{align*}
Note that this choice amounts to the destinations quantizing their
observations with the quantization noise level set to the power level at
which their own {\sf private}-codewords ({\em e.g.}, $Z_1$ in the case of
destination~3) is received. This is consistent with the intuition that the
information they forward on to the other destination is used to recover
only the {\sf public}-codewords.

With these choices, it can be shown that
Theorem~\ref{thm:destcoopgenericschemes} implies that the non-negative
rates $r_{X_1},r_{X_2},r_{U_1},r_{U_2},r_3,r_4$ can be achieved if the
following conditions are satisfied 
\begin{align*}
r_{X_1}&\leq [n_{1,3}-n_{1,4}]_+ -2,\\
r_{U_1}&\leq \max\left( n_{1,3} - [n_{1,3}-n_{1,4}]_+, 
                        n_{1,4} - [n_{2,4}-n_{2,3}]_+ \right) -\log 36,\\
r_{U_1}&\leq n_{1,3} - [n_{1,3}-n_{1,4}]_+ + r_4 - 3,\\
r_{U_2}&\leq \max\left( n_{2,3} - [n_{1,3}-n_{1,4}]_+, 
                        n_{2,4} - [n_{2,4}-n_{2,3}]_+ \right) -\log 36,\\
r_{U_2}&\leq  [n_{2,3} - [n_{1,3}-n_{1,4}]_+]_+ + r_4 -3,\\
r_4 &\leq [n_C - [n_{1,3}-n_{1,4}]_+]_+ - 1,\\
r_{U_1} + r_{U_2} &\leq 
\log\Bigg( 1 + 
\left|\frac{g_{1,3}}{\alpha_1}\right|^2+\left|\frac{g_{2,4}}{\alpha_2}\right|^2+
 \left|\frac{g_{1,4}}{\alpha_2}\right|^2+\left|\frac{g_{2,3}}{\alpha_1}\right|^2 \\
 &\qquad\qquad\qquad\qquad\qquad
 +\left|\frac{g_{1,3}g_{2,4}}{\alpha_1\alpha_2}\right|^2 +
 \left|\frac{g_{1,4}g_{2,3}}{\alpha_1\alpha_2}\right|^2 -
2\left|\frac{g_{1,3}g_{2,4}g_{1,4}g_{2,3}}{\alpha_1^2\alpha_2^2}\right|\cos\theta \Bigg) -\log 36,\\
r_{U_1} + r_{U_2} &\leq \max( n_{1,3},n_{2,3} ) -
                        [n_{1,3}-n_{1,4}]_+ + r_4 -3,
\end{align*}
\begin{align*}
r_{U_1} + r_4  &\leq \max\bigg( \max(n_{1,3},n_C) - [n_{1,3}-n_{1,4}]_+,\\
&\qquad\qquad\qquad
 [n_{1,4}-[n_{2,4}-n_{2,3}]_+]_+ + [n_C - [n_{1,3}-n_{1,4}]_+]_+\bigg)-\log 36,\\
r_{U_1} + r_4  &\leq n_{1,3} - [n_{1,3}-n_{1,4}]_+ + r_4 -3,\\
r_{U_2} + r_4  &\leq \max\bigg( \max(n_{2,3},n_C) - [n_{1,3}-n_{1,4}]_+,\\
&\qquad\qquad\qquad
 (n_{2,4}-[n_{2,4}-n_{2,3}]_+) + [n_C - [n_{1,3}-n_{1,4}]_+]_+\bigg)-\log 36,\\
r_{U_2} + r_4  &\leq [n_{2,3} - [n_{1,3}-n_{1,4}]_+]_+ + r_4 -3,\\
r_{U_2} + r_4 + r_{U_1} &\leq \log\Bigg( 1 + 
\left|\frac{g_{1,3}}{\alpha_1}\right|^2+\left|\frac{g_{2,4}}{\alpha_2}\right|^2+
 \left|\frac{g_{1,4}}{\alpha_2}\right|^2+\left|\frac{g_{2,3}}{\alpha_1}\right|^2
+\left|\frac{g_C}{\alpha_1}\right|^2 + \left|\frac{g_C g_{1,4}}{\alpha_1 \alpha_2}\right|^2 + \left|\frac{g_C g_{2,4}}{\alpha_1 \alpha_2}\right|^2\\
 &\qquad\qquad\quad +
 \left|\frac{g_{1,3}g_{2,4}}{\alpha_1\alpha_2}\right|^2 +
 \left|\frac{g_{1,4}g_{2,3}}{\alpha_1\alpha_2}\right|^2 -
 2\left|\frac{g_{1,3}g_{2,4}g_{1,4}g_{2,3}}{\alpha_1^2\alpha_2^2}\right|\cos\theta \Bigg) -\log 36,\text{ and}\\
r_{U_2} + r_4 + r_{U_1} &\leq \max( n_{1,3},n_{2,3},n_C ) - 
                               [n_{1,3}-n_{1,4}]_+ + r_4 - 3,
\end{align*}
and the corresponding inequalities with subscripts 1 and 2 exchanged,
and 3 replaced with 4. Above, we used
\begin{align*}
\alpha_1 &=
 \sqrt{\max\left(1,\frac{\max(1,|g_{1,3}|^2)}{\max(1,|g_{1,4}|^2)}\right)},
 \text{ and}\\
\alpha_2 &=
 \sqrt{\max\left(1,\frac{\max(1,|g_{2,4}|^2)}{\max(1,|g_{2,3}|^2)}\right)}.
\end{align*}
To illustrate, we show how a couple of the above conditions are arrived at.
The rest are also derived similarly. Two of the conditions on $r_{U_1}+
r_{U_2}$ from Theorem~\ref{thm:destcoopgenericschemes} are
\begin{align}
 r_{U_1}+r_{U_2} &\leq I(U_1,U_2;Y_3,V_4|X_3,X_4), \text{ and} 
\label{eq:u1u21} \\
 r_{U_1}+r_{U_2} &\leq I(U_1,U_2;Y_3|X_3,X_4) + r_4 -
I(Y_4;V_4|X_3,X_4,U_1,U_2,Y_3). \label{eq:u1u22}
\end{align}
Below, we show the following:
\begin{align*}
I(U_1,U_2;Y_3,V_4|X_3,X_4) &\geq \log\Bigg( 1 + 
\left|\frac{g_{1,3}}{\alpha_1}\right|^2+\left|\frac{g_{2,4}}{\alpha_2}\right|^2+
 \left|\frac{g_{1,4}}{\alpha_2}\right|^2+\left|\frac{g_{2,3}}{\alpha_1}\right|^2 \\
 &\qquad\qquad\quad
 +\left|\frac{g_{1,3}g_{2,4}}{\alpha_1\alpha_2}\right|^2 +
 \left|\frac{g_{1,4}g_{2,3}}{\alpha_1\alpha_2}\right|^2 -
2\left|\frac{g_{1,3}g_{2,4}g_{1,4}g_{2,3}}{\alpha_1^2\alpha_2^2}\right|\cos\theta
\Bigg) -\log 36,\\
I(U_1,U_2;Y_3|X_3,X_4) &\geq \max( n_{1,3},n_{2,3} ) -
                        [n_{1,3}-n_{1,4}]_+ - 2,\text{ and}\\
I(Y_4;V_4|X_3,X_4,U_1,U_2,Y_3) &\leq 1.
\end{align*}
This will allow us to conclude that in order for $r_{U_1}$, $r_{U_2}$ to
satisfy \eqref{eq:u1u21}-\eqref{eq:u1u22}, it is enough if they satisfy
\begin{align*}
r_{U_1} + r_{U_2} &\leq 
\log\Bigg( 1 + 
\left|\frac{g_{1,3}}{\alpha_1}\right|^2+\left|\frac{g_{2,4}}{\alpha_2}\right|^2+
 \left|\frac{g_{1,4}}{\alpha_2}\right|^2+\left|\frac{g_{2,3}}{\alpha_1}\right|^2 \\
 &\qquad\qquad\qquad\qquad\qquad
 +\left|\frac{g_{1,3}g_{2,4}}{\alpha_1\alpha_2}\right|^2 +
 \left|\frac{g_{1,4}g_{2,3}}{\alpha_1\alpha_2}\right|^2 -
2\left|\frac{g_{1,3}g_{2,4}g_{1,4}g_{2,3}}{\alpha_1^2\alpha_2^2}\right|\cos\theta \Bigg) -\log 36,\text{ and}\\
r_{U_1} + r_{U_2} &\leq \max( n_{1,3},n_{2,3} ) -
                        [n_{1,3}-n_{1,4}]_+ + r_4 -3.
\end{align*}

From the choices for the auxiliary random variables we made,
\begin{align*}
I(U_1,U_2;Y_3,V_4|X_3,X_4) &=
I\left(\left[\begin{array}{c}U_1\\U_2\end{array}\right];
  \left[\begin{array}{c}\frac{Y_3}{\alpha_1}\\
       \frac{V_4}{\alpha_2}\end{array}\right]\right)\\
 &= I\left(\left[\begin{array}{c}U_1\\U_2\end{array}\right];
   \left[\begin{array}{cc}g_{1,3}&g_{2,3}\\g_{1,4}&g_{2,4}\end{array}\right]
   \left[\begin{array}{c}U_1\\U_2\end{array}\right] +
   \left[\begin{array}{c} \frac{g_{1,3}Z_1+g_{2,3}Z_2+N_3}{\alpha_1}\\
      \frac{g_{1,4}Z_1+g_{2,4}Z_2+N_4}{\alpha_2}\end{array}\right] + 
   \left[\begin{array}{c}0\\\frac{Q_4}{\alpha_2}\end{array}\right]\right)\\
 &\geq I\left(\left[\begin{array}{c}U_1\\U_2\end{array}\right];
   \left[\begin{array}{cc}g_{1,3}&g_{2,3}\\g_{1,4}&g_{2,4}\end{array}\right]
   \left[\begin{array}{c}U_1\\U_2\end{array}\right] +
   \left[\begin{array}{c} \frac{g_{1,3}Z_1+g_{2,3}Z_2+N_3}{\alpha_1}\\
      \frac{g_{1,4}Z_1+g_{2,4}Z_2+N_4}{\alpha_2}\end{array}\right] + 
   \left[\begin{array}{c} {Q'_3}\\
          \frac{Q_4}{\alpha_2}\end{array}\right]\right)\\
 &= I\left({\bf U};{\bf H}{\bf U} + {\bf N} + {\bf Q}\right),
\end{align*}
where $Q'_3$ is a unit variance Gaussian random variable independent of
everything else. In the last step, we defined the Gaussian vectors ${\bf U}, 
{\bf H}, {\bf N},$ and ${\bf Q}$. Note that $\frac{Q_4}{\alpha_2}$ is a
unit variance Gaussian random variable which makes ${\bf Q}$ a Gaussian
random vector whose covariance matrix is the identity matrix ${\bf I}$.
Also, note that the covariance matrix of ${\bf U}$ is ${\bf K}_{\bf U} =
 K^2 {\bf I}$. Continuing,
\begin{align*}
I(U_1,U_2;Y_3,V_4|X_3,X_4) 
  &\geq \log \frac{\left| {\bf H}{\bf K}_{\bf U}{\bf H}^\dagger + 
        {\bf K}_{\bf N} + {\bf I} \right|}{\left| {\bf K}_{\bf N} +
         {\bf I}\right|}\\
  &\geq \log \frac{\left| {\bf H}{\bf K}_{\bf U}{\bf H}^\dagger + 
         {\bf I} \right|}{\left| {\bf K}_{\bf N} + {\bf I}\right|}
\end{align*}
From the choices made for the variances of $Z_1,Z_2$, we can find a uniform
upperbound for the denominator for all possible channels:
\[ \left| {\bf K}_{\bf N} + {\bf I}\right| \leq 9.\]
Evaluating the lowerbound on $I(U_1,U_2;Y_3,V_4|X_3,X_4)$ using this and
substituting $K=1/2$, we can show that
\begin{align*}
I(U_1,U_2;Y_3,V_4|X_3,X_4) &\geq \log\Bigg( 1 + 
\left|\frac{g_{1,3}}{\alpha_1}\right|^2+\left|\frac{g_{2,4}}{\alpha_2}\right|^2+
 \left|\frac{g_{1,4}}{\alpha_2}\right|^2+\left|\frac{g_{2,3}}{\alpha_1}\right|^2 \\
 &\qquad\qquad\quad
 +\left|\frac{g_{1,3}g_{2,4}}{\alpha_1\alpha_2}\right|^2 +
 \left|\frac{g_{1,4}g_{2,3}}{\alpha_1\alpha_2}\right|^2 -
2\left|\frac{g_{1,3}g_{2,4}g_{1,4}g_{2,3}}{\alpha_1^2\alpha_2^2}\right|
  \cos\theta \Bigg) -\log 36.
\end{align*}
Similarly,
\begin{align*}
I(U_1,U_2;Y_3|X_3,X_4) &=
I(U_1,U_2;g_{1,3}U_1+g_{2,3}U_2+g_{1,3}Z_1+g_{2,3}Z_2+N_3)\\
&\geq \log\left( 1 + \frac{|g_{1,3}|^2/K + |g_{2,3}|^2/K}
 {1+\max\left(1,\frac{\max(1,|g_{1,3}|^2)}{\max(1,|g_{1,4}|^2)}\right)}
\right)\\
&\geq \max( n_{1,3},n_{2,3} ) - [n_{1,3}-n_{1,4}]_+ - \log 4,\\
\intertext{and}
I(Y_4;V_4|X_3,X_4,U_1,U_2,Y_3) 
 &\leq I(g_{1,4}Z_1 + g_{2,4}Z_2 + N_4; 
        g_{1,4}Z_1 + g_{2,4}Z_2 + N_4 + Q_4)\\
 &= \log\left( 1 + \frac{\alpha_2^2}{1+\alpha_2^2}\right)
 \leq \log 2 = 1.
\end{align*}

Applying Fourier-Motzkin elimination to the set of all conditions on the
rates, we can show that a sum-rate equal to the minimum of the following
terms is achievable with a gap of at most 15 bits in this regime.
\begin{align*}
u'_2&=
\max(n_{2,4},n_{2,3}) + \left(\max(n_{1,3},n_{2,3},n_C)-n_{2,3}\right),\\
u'_3&=
\max(n_{1,3},n_{1,4}) + \left(\max(n_{2,4},n_{1,4},n_C)-n_{1,4}\right),\\
u'_4&=\max(n_{1,3},n_C)+\max(n_{2,4},n_C),\text{ and}\\
u'_5&=\log\bigg( 1 + 
 \left(|g_{1,3}|^2+|g_{2,4}|^2+|g_{1,4}|^2+|g_{2,3}|^2\right)\notag\\
&\qquad\qquad\qquad\qquad\quad +
 \left( |g_{1,3}g_{2,4}|^2 + |g_{1,4}g_{2,3}|^2 -
2|g_{1,3}g_{2,4}g_{1,4}g_{2,3}|\cos\theta\right) \bigg).
\end{align*}

\section{Proof of the upperbounds of Theorems~\ref{thm:destcoopLD} and \ref{thm:destcoopG}}\label{app:upperbounds}

Upperbounds 1-3 are new, upperbounds 4 and 5 are cut-set upperbounds which
also appeared in the two-user interference channel with source
cooperation~\cite{SourceCoop}. Below, we prove upperbounds 1-3 and, for
completeness, repeat the proofs for upperbound 4 and 5.

\noindent{\em Upperbound 1:}

We create two dummy channels in both of which, all the noise processes are
independent of those in the original channel, but have identical
distributions to their counterparts in the original channel. All the nodes
use the same strategies as in the original problem ({\em i.e.}, same
codebooks at the nodes 1 and 2, and the same $f_{k,t}$'s at nodes 3 and 4),
but the messages transmitted by the nodes are different from that in the
original channel as explained below. In the first dummy channel (where all
quantities are denoted by adding a prime $\prime$), the message at node 1
is identical to the message at node 1 in the original channel, {\em i.e.},
$M_1^\prime=M_1$, however, the message $M_2^\prime$ at node 2 is
independent of the messages $M_1,M_2$ and distributed uniformly over its
alphabet. We note that $X_1^\prime=X_1$, but $X_2^\prime, X_3^\prime$, and
$X_4^\prime$ are, in general, different from their counterparts in the
original channel. Similarly, $M_2^{\prime\prime}=M_2$, but
$M_1^{\prime\prime}$ is independent of $M_1,M_2,M_2^\prime$ and distributed
uniformly over its alphabet. We start from Fano's inequality.
\begin{align*}
T(R_1+R_2&-o(\epsilon))\\
&\leq I(M_1;Y_3^T)+I(M_2;Y_4^T)\\
&\leq I(M_1;Y_3^T,g_{1,4}^\ast(X_1^{\prime T})+g_{3,4}(X_3^{\prime T}),M_2^\prime) +
I(M_2;Y_4^T,g_{2,3}^\ast(X_2^{\prime\prime T})+g_{4,3}(X_4^{\prime\prime T}),M_1^{\prime\prime})\\
&\leq I(M_1;Y_3^T,g_{1,4}^\ast(X_1^{\prime T})+g_{3,4}(X_3^{\prime T})|M_2^\prime) +
I(M_2;Y_4^T,g_{2,3}^\ast(X_2^{\prime\prime T})+g_{4,3}(X_4^{\prime\prime T})|M_1^{\prime\prime}).
\end{align*}
These two symmetric terms can be further upperbounded. Below we will show
how the first is upperbounded; the second term can be similarly upperbounded.
\begin{align*}
I(M_1;Y_3^T&,g_{1,4}^\ast(X_1^{\prime T})+g_{3,4}(X_3^{\prime T})|M_2^\prime)\\
&=H(g_{1,4}^\ast(X_1^{\prime T})+g_{3,4}(X_3^{\prime T})|M_2^\prime) + H(Y_3^T|g_{1,4}^\ast(X_1^{\prime T})+g_{3,4}(X_3^{\prime T}),M_2^\prime)\\
&\qquad-H(Y_3^T|M_1,M_2^\prime) - H(g_{1,4}^\ast(X_1^{\prime T})+g_{3,4}(X_3^{\prime T})|Y_3^T,M_1,M_2^\prime)\\
&\stackrel{\text{(a)}}{=} H(g_{1,4}^\ast(X_1^{\prime T})+g_{3,4}(X_3^{\prime T})|M_2^\prime)
 + H(Y_3^T|g_{1,4}^\ast(X_1^{\prime T})+g_{3,4}(X_3^{\prime T}),M_2^\prime)\\
&\qquad - H(g_{2,3}^\ast(X_2^T)+g_{4,3}(X_4^T)|M_1,M_2^\prime) - H(g_{1,4}^\ast(X_1^{\prime T})+g_{3,4}(X_3^{\prime T})|Y_3^T,M_1,M_2^\prime)\\
&\stackrel{\text{(b)}}{=} H(g_{1,4}^\ast(X_1^{\prime T})+g_{3,4}(X_3^{\prime T})|M_2^\prime)
 + H(Y_3^T|g_{1,4}^\ast(X_1^{\prime T})+g_{3,4}(X_3^{\prime T}),M_2^\prime)\\
 &\qquad- H(g_{2,3}^\ast(X_2^T)+g_{4,3}(X_4^T)|M_1) - H(g_{1,4}^\ast(X_1^{\prime T})+g_{3,4}(X_3^{\prime T})|Y_3^T,M_1,M_2^\prime)\\
&\stackrel{\text{(c)}}{=} H(g_{1,4}^\ast(X_1^{\prime T})+g_{3,4}(X_3^{\prime T})|M_2^\prime)
 + H(Y_3^T|g_{1,4}^\ast(X_1^{\prime T})+g_{3,4}(X_3^{\prime T}),M_2^\prime)\\
 &\qquad - H(g_{2,3}^\ast(X_2^T)+g_{4,3}(X_4^T)|M_1) - H(g_{1,4}^\ast(X_1^{\prime T})+g_{3,4}(X_3^{\prime T})|M_1,M_2^\prime).
\end{align*}
where (a) follows from the fact that $Y_3[t]=g_{1,3}(X_1[t])+g_{2,3}^\ast(X_2[t])+g_{4,3}(X_4[t])$ and 
$g_{1,3}(X_1[t])$ is a deterministic function of $M_1$, and (b) is due to the
independence of\\ $(M_1,g_{2,3}^\ast(X_2^T),g_{4,3}(X_4^T))$ and $M_2^\prime$. Equality (c) follows
from the fact that conditioned on $M_1$, the primed quantities and the unprimed quantities are independent. Further, we can upperbound the second and fourth
terms as follows
\begin{align*}
H(Y_3^n|g_{1,4}^\ast(X_1^{\prime T})+g_{3,4}(X_3^{\prime T})&, M_2^\prime) \leq 
 H(Y_3^n|g_{1,4}^\ast(X_1^{\prime T})+g_{3,4}(X_3^{\prime T})),
\end{align*}
\begin{align*}
-&H(g_{1,4}^\ast(X_1^{\prime T})+g_{3,4}(X_3^{\prime T})|M_1,M_2^\prime) \\
 &= \sum_{t=1}^T -H(g_{1,4}^\ast(X_1^\prime[t])+g_{3,4}(X_3^\prime[t])|g_{1,4}^\ast(X_1^{\prime t-1})+
                      g_{3,4}(X_3^{\prime t-1}),M_1,M_2^\prime)\\
 &\leq \sum_{t=1}^T -H(g_{1,4}^\ast(X_1^\prime[t])+g_{3,4}(X_3^\prime[t])|g_{1,4}^\ast(X_1^{\prime t-1})+g_{3,4}(X_3^{\prime t-1}),M_1,M_2^\prime,X_1^\prime[t],g_{3,4}(X_3^\prime[t]))\\
 &= \sum_{t=1}^T -H(g_{1,4}^\ast(X_1^\prime[t])|X_1^\prime[t]).
\end{align*}
We combine these and use the following facts
\begin{align*}
H(g_{1,4}^\ast(X_1^{\prime T})+g_{3,4}(X_3^{\prime T})|M_1^\prime) &= H(g_{1,4}^\ast(X_1^T)+g_{3,4}(X_3^T)|M_1),\\
H(g_{2,3}^\ast(X_2^{\prime\prime T})+g_{4,3}(X_4^{\prime\prime T})|M_2^{\prime\prime})
  &= H(g_{2,3}^\ast(X_2^T)+g_{4,3}(X_4^T)|M_2).
\end{align*}
We arrive at
\begin{align*}
T(R_1+R_2-o(\epsilon)) &\leq 
H(Y_3^T|g_{1,4}^\ast(X_1^{\prime T})+g_{3,4}(X_3^{\prime T})) +H(Y_4^T|g_{2,3}^\ast(X_2^{\prime\prime T})+g_{4,3}(X_4^{\prime\prime T}))\\
&\qquad
- \left(\sum_{t=1}^T  H(g_{1,4}^\ast(X_1^\prime[t])|X_1^\prime[t]) 
   + H(g_{2,3}^\ast(X_2^{\prime\prime}[t])|X_2^{\prime\prime}[t]) \right).
\end{align*}

\noindent{\em Linear deterministic case:}\\
We have
\begin{align*}
R_1+R_2-o(\epsilon) &\leq \max(n_{2,3},n_{1,3}-n_{1,4}+n_{4,3},n_{4,3})-0 +
\max(n_{1,4},n_{2,4}-n_{2,3}+n_{3,4},n_{3,4})-0\\
&= \max(n_{2,3},n_{1,3}-n_{1,4}+n_C,n_C) + \max(n_{1,4},n_{2,4}-n_{2,3}+n_C,n_C).
\end{align*}
 
\noindent{\em Gaussian case:}\\
We have,
\begin{align*}
R_1+R_2-o(\epsilon) 
  \leq&\quad \left\{\begin{array}{ll}
             \log\left(1 +\left(|g_{2,3}|+|g_C|+
              \left|\frac{g_{1,3}g_C}{g_{1,4}}\right|\right)^2
              +\left|\frac{g_{1,3}}{g_{1,4}}\right|^2\right), 
                &\text{ if } |g_{1,4}|>\max(1,|g_C|)\\
             \log\left(1 +\left(|g_{2,3}|+|g_C|+
              |g_{1,3}|\right)^2\right), &\text{ otherwise}
        \end{array}\right. \\  
    &\quad +   \left\{\begin{array}{ll}
              \log\left(1 +\left(|g_{1,4}|+|g_C|+
              \left|\frac{g_{2,4}g_C}{g_{2,3}}\right|\right)^2
              +\left|\frac{g_{2,4}}{g_{2,3}}\right|^2\right),
                &\text{ if } |g_{2,3}|>\max(1,|g_C|)\\
             \log\left(1 +\left(|g_{1,4}|+|g_C|+
               |g_{2,4}|\right)^2\right), &\text{ otherwise}
        \end{array}\right.\\
\end{align*}

\noindent{\em Upperbounds 2 and 3:}

We start from Fano's inequality.
\begin{align*}
T(R_1+R_2-o(\epsilon))&\leq I(M_1;Y_3^T)+I(M_2;Y_4^T)\\
&\leq I(M_1;Y_3^T) + I(M_2;Y_4^T,g_{2,3}^\ast(X_2^T),M_1)\\
&\leq I(M_1;Y_3^T) + I(M_2;Y_4^T,g_{2,3}^\ast(X_2^T),|M_1)\\
&\leq \sum_{t=1}^T I(M_1;Y_3[t]|Y_3^{t-1}) + 
            I(M_2;Y_4[t],g_{2,3}^\ast(X_2[t]),|Y_4^{t-1},g_{2,3}^\ast(X_2^{t-1}),M_1).
\end{align*}
Below, we upperbound these terms separately.
\begin{align*}
I(M_1;Y_3[t]|Y_3^{t-1})&=H(Y_3[t]|Y_3^{t-1}) - H(Y_3[t]|Y_3^{t-1},M_1)\\
&\leq H(Y_3[t]) - H(Y_3[t]|Y_3^{t-1},Y_4^{t-1},M_1)\\
&\stackrel{\text{(a)}}{\leq} H(Y_3[t]) - H(g_{2,3}^\ast(X_2[t])|Y_4^{t-1},g_{2,3}^\ast(X_2^{t-1}),M_1),
\end{align*}
where (a) follows from the fact that $Y_3[t]=g_{1,3}(X_1[t])+g_{2,3}^\ast(X_2[t])+g_{4,3}(X_4[t])$, and 
$X_1[t]$ is a deterministic function of $M_1$, $X_4[t]$ is a deterministic
function of $Y_4^{t-1}$, and $g_{1,3}$ and $g_{4,3}$ are deterministic maps.
\begin{align*}
I(M_2;Y_4[t],&g_{2,3}^\ast(X_2[t])|Y_4^{t-1},g_{2,3}^\ast(X_2^{t-1}),M_1)\\
&=I(M_2;g_{2,3}^\ast(X_2[t])|Y_4^{t-1},g_{2,3}^\ast(X_2^{t-1}),M_1)
+I(M_2;Y_4[t]|Y_4^{t-1},g_{2,3}^\ast(X_2^t),M_1).
\end{align*}
We upperbound these two terms separately now.
\begin{align*}
I(M_2;g_{2,3}^\ast(X_2[t])|&Y_4^{t-1},g_{2,3}^\ast(X_2^{t-1}),M_1)\\
&=H(g_{2,3}^\ast(X_2[t])|Y_4^{t-1},g_{2,3}^\ast(X_2^{t-1}),M_1) - H(g_{2,3}^\ast(X_2[t])|X_2[t]),
\end{align*}
which follows from the channel model (memorylessness and independence of
the noise processes at the different nodes).
\begin{align*}
I(M_2&;Y_4[t]|Y_4^{t-1},g_{2,3}^\ast(X_2^t),M_1)\\
&\stackrel{\text{(a)}}{=} I(M_2;Y_4[t]|Y_4^{t-1},g_{2,3}^\ast(X_2^t),M_1,X_3[t])\\
&= H(Y_4[t]|Y_4^{t-1},g_{2,3}^\ast(X_2^t),M_1,X_3[t]) - H(Y_4[t]|Y_4^{t-1},g_{2,3}^\ast(X_2^t),X_3[t],M_1,M_2)\\
&\leq H(X_2[t]+g_{1,4}^\ast(X_1[t])|g_{2,3}^\ast(X_2[t]),X_1[t]) - H(Y_4[t]|Y_4^{t-1},g_{2,3}^\ast(X_2^t),X_3[t],M_1,M_2)\\
&\stackrel{\text{(b)}}{=} H(X_2[t]+g_{1,4}^\ast(X_1[t])|g_{2,3}^\ast(X_2[t]),X_1[t]) - H(g_{1,4}^\ast(X_1[t])|X_1[t]),
\end{align*}
where (a) can be seen by noting that: (1) $X_3[t]$ is a deterministic function
$f_{3,t}$ of $Y_3^{t-1}$, (2) $Y_3^{t-1}$ in turn is such that
\begin{align*}
Y_3(s)=g_{1,3}(X_1(s))+g_{2,3}^\ast(X_2(s))+g_{4,3}(X_4(s)),\; s\leq t-1,
\end{align*}
and (3) for all $s\leq t-1$, $g_{1,3}(X_1(s))$ is a deterministic function of $M_1$,
and $g_{4,3}(X_4(s))$ is a deterministic function of $Y_4^{s-1}$. Also, (b) follows
from the channel model (memorylessness and the independence of the noise
processes at the different nodes) and the fact that $X_1[t]$ is a
deterministic function $M_1$.

\noindent Combining everything, we have
\begin{align*}
T(R_1+R_2-o(\epsilon)) &\leq 
\sum_{t=1}^T \left\{ H(Y_3[t]) 
-H(g_{2,3}^\ast(X_2[t])|X_2[t]) \right\}\\
&\quad\qquad +\left\{ H(X_2[t]+g_{1,4}^\ast(X_1[t])|g_{2,3}^\ast(X_2[t]),X_1[t]) - H(g_{1,4}^\ast(X_1[t])|X_1[t]) \right\}
\end{align*}

\noindent{\em Linear deterministic channel:}
We have
\begin{align*}
R_1+R_2-o(\epsilon) &\leq \{\max(n_{1,3},n_{2,3},n_{4,3})-0\} + \{[n_{2,4}-n_{2,3}]_+-0\}\\
&= \max(n_{1,3},n_{2,3},n_C) + [n_{2,4}-n_{2,3}]_+.
\end{align*}

\noindent{\em Gaussian channel:}
We have
\begin{align*}
R_1+R_2-o(\epsilon) &\leq \log(1+(|g_{1,3}|+|g_{2,3}|+|g_C|)^2) +
\log\left(1+\frac{|g_{2,4}|^2}{\max(1,|g_{2,3}|^2)}\right).
\end{align*}

\noindent{\em Upperbound 4:}

This is a simple cut-set upperbound~\cite{CoverThomas} with nodes 1 and 4
on one side of the cut and nodes 2 and 3 on the other. It is easy to verify
that
\begin{align*}
 R_1 &\leq \max_{p_{X_1}} I(X_1;Y_3,Y_2),\\
 R_2 &\leq \max_{p_{X_2}} I(X_2;Y_4,Y_1).
\end{align*}
Under the linear deterministic model, this translates to an upperbound on
the sum-rate of
\[ R_1 + R_2 \leq \max(n_{1,3},n_C) + \max(n_{2,4},n_C),\]
and for the Gaussian case, we get an upperbound of
\[ R_1 + R_2 \leq \log\left( 1 + |g_{1,3}|^2 + |g_C|^2\right) + \log\left(
1 + |g_{2,4}|^2 + |g_C|^2\right).\]

\noindent{\em Upperbound 5:}

This is also a simple cut-set upperbound. Nodes 1 and 2 are on one side of
the cut and nodes 3 and 4 are on the other. The resulting upperbound on the
sum-rate is
\begin{align*}
 R_1 + R_2 &\leq \max_{p_{X_1,X_2}} I(X_1;X_2; Y_3,Y_4).
\end{align*}
For the linear deterministic case, this gives
\[R_1 + R_2 \leq  \left\{\begin{array}{ll}
\max(n_{1,3}+n_{2,4},n_{1,4}+n_{2,3}),
 &\text{ if }n_{1,3}-n_{2,3}\neq n_{1,4}-n_{2,4},\\
\max(n_{1,3},n_{2,4},n_{1,4},n_{2,3}),
 &\text{ otherwise},
\end{array}\right.
\]
and for the Gaussian case, using the fact the eigenvalues of the input
($[X_1, X_2]$) covariance matrix cannot exceed 2, we may upperbound the
sum-rate by
\begin{align*}
 R_1 + R_2 \leq \log\bigg( 1 &+
2 \left(|g_{1,3}|^2+|g_{2,4}|^2+|g_{1,4}|^2+|g_{2,3}|^2\right)\\ &+
4 \left( |g_{1,3}g_{2,4}|^2 + |g_{1,4}g_{2,3}|^2 -
2|g_{1,3}g_{2,4}g_{1,4}g_{2,3}|\cos\theta\right) \bigg).
\end{align*}

\end{document}